\documentclass[aps,prd,reprint,groupaddress,noshowpacs,noshowkeys]{revtex4-1}

\usepackage{amsmath,amssymb,mathrsfs}

\usepackage{graphicx}% Include figure files
\usepackage{dcolumn}% Align table columns on decimal point
\usepackage{bm}% bold math
\usepackage{xcolor}
\usepackage[pdftex,
  hidelinks,
  colorlinks=true,
  linkcolor=black,
  anchorcolor=black,
  citecolor=black,
  urlcolor=black,
  pdfauthor= {Michael Glinsky},
  pdfsubject = {MST for MHD},
  pdfdisplaydoctitle,
  bookmarks=true,
  bookmarksopen=true]{hyperref}
  
\begin{document}
\title{Mallat Scattering Transformation based surrogate for MagnetoHydroDynamics}

\author{Michael E. Glinsky}
\affiliation{qiTech Consulting, Santa Fe, NM, USA}

\author{Kathryn Maupin}
\affiliation{Sandia National Laboratories, Albuquerque, NM, USA}

%\date{\today, arXiv v1, Computational Mechanics rev2, version 8}

%\pacs{11.10.Gh}

\begin{abstract}
A Machine and Deep Learning (MLDL) methodology is developed and applied to give a high fidelity, fast surrogate for 2D resistive MagnetoHydroDynamic (MHD) simulations of Magnetic Liner Inertial Fusion (MagLIF) implosions.  The resistive MHD code \texttt{GORGON} is used to generate an ensemble of implosions with different liner aspect ratios, initial gas preheat temperatures (that is, different adiabats), and different liner perturbations.  The liner density and magnetic field as functions of $x$, $y$, and $t$ were generated.  The Mallat Scattering Transformation (MST) is taken of the logarithm of both fields and a Principal Components Analysis (PCA) is done on the logarithm of the MST of both fields.  The fields are projected onto the PCA vectors and a small number of these PCA vector components are kept.  Singular Value Decompositions of the cross correlation of the input parameters to the output logarithm of the MST of the fields, and of the cross correlation of the SVD vector components to the PCA vector components are done.  This allows the identification of the PCA vectors vis-a-vis the input parameters.  Finally, a Multi Layer Perceptron (MLP) neural network with ReLU activation and a simple three layer encoder/decoder architecture is trained on this dataset to predict the PCA vector components of the fields as a function of time.  Details of the implosion, stagnation, and the disassembly are well captured.  Examination of the PCA vectors and a permutation importance analysis of the MLP show definitive evidence of an inverse turbulent cascade into a dipole emergent behavior.  The orientation of the dipole is set by the initial liner perturbation.  The analysis is repeated with a version of the MST which includes phase, called Wavelet Phase Harmonics (WPH).  While WPH do not give the physical insight of the MST, they can and are inverted to give field configurations as a function of time, including field-to-field correlations.
\end{abstract}

\maketitle

\section{Introduction}
The major challenge for physics-based machine learning is to replace expensive finite element and finite volume computer simulations with fast Machine and Deep Learning (MLDL) based surrogates that reproduce all the structure and emergent behaviors of the system.  These surrogates can then be used for experimental design and analysis.  Fundamentally, they can be used for hypothesis testing, theoretical model verification, and model extrapolation and scaling.  The challenge has been to capture the rich texture of physical systems, then to reproduce them, not only to predict the texture of one field, but to predict the rich correlations between the fields.  Attempts using both Gaussian process simulation \citep{rasmussen2006} and convolutional network techniques combined with reduced order models \citep{goodfellow2016deep, LeCun1989} have had modest success.  Inspired by the success of the Mallat Scattering Transformation (MST) \citep{Mallat2012,Bruna2013} and the Wavelet Phase Harmonics (WPH) enhancement of the MST \citep{mallat2020phase}, which includes phase, in classifying and reproducing textures of physical systems \citep{zhang2021maximum}, this paper presents a simple MLDL methodology based on the MST and WPH to give a high fidelity, fast surrogate of 2D resistive MagnetoHydroDynamics (MHD).

A major goal of this development is not only to reproduce the system evolution, but to do it in a way that can lead to physical insight.  Much of the work to date in physics-based MLDL has approached the challenge using MLDL as a black-box of ingredients to be combined, with success judged by final performance metrics.  Attributes produced in the analysis are abstract vectors with little or no physical meaning.  Displays of those vectors are rarely given.  The methodology presented in this paper is inspired by fundamental understanding of the physics; has physical interpretations of the attributes, vectors, and MLDL structures; and has displays of the attributes, vectors, and MLDL performance that leads to important physical insights of the nonlinear and quantum dynamics of the system.

The system and geometry that is chosen to prototype this MLDL methodology is Magnetized Liner Inertial Fusion (MagLIF). MagLIF is a magneto-inertial fusion concept currently being explored at Sandia's Z Pulsed Power Facility \citep{Slutz2010,Awe2013,Gomez2014,McBride2012}.  MagLIF produces thermonuclear fusion conditions by driving mega-amps of current through a low-Z conducting liner. The subsequent implosion of the liner containing a preheated and pre-magnetized fuel of deuterium or deuterium-tritium compresses and heats the system, creating a plasma with fusion relevant conditions.  Axial symmetry is assumed and the dynamics is simulated in the Cartesian, perpendicular plane.

It is well-known that the large accelerations of the liner, as it drives the implosion of the gas, cause the liner to go linearly unstable to the Magnetic Rayleigh Taylor (MRT) instability \citep{Seyler2018}.  During the implosion, it has sufficient time to evolve well into to the nonlinear regime.  Increasing the AR will increase the linear growth rate by increasing the acceleration of the liner.  Reducing the preheat temperature $T_0$ will put the implosion on a higher adiabat, allowing the implosion to reach higher compression ratios and a smaller radius at stagnation.  This gives the instability more time to evolve, and puts a larger demand on the uniformity of the imploding surface.  For laser-driven indirect ICF capsules, without a large applied magnetic field, the Rayleigh-Taylor instability exhibits a normal turbulent cascade that destroys the implosion well before it reaches the required compression, if allowed to grow.  For this reason, the implosions need to have a larger AR, a larger $T_0$ (be put on a lower adiabat), and to have very small perturbations to the capsule surface.  Unfortunately, this means that it has been very difficult to reach the needed conditions at stagnation for thermonuclear burn.  There is indication that this is not the case for MagLIF.  A double helical structure is observed by \citet{Awe2013} with helical threads $30 \, \mu m$ in diameter, separated by $100 \, \mu m$ reaching very high compression ratios.  Also, evidence of an inverse turbulent cascade in the liner structure has been seen by \citet{yager2018} on ultra-thin foils driven at less than 1 MA.   This is the reason that AR and $T_0$, along with how the MRT is seeded, are the parameters that are varied.  The scientific goal of this study is to characterize and understand the nonlinear evolution of the MRT.

The finite volume, parallel, resistive MHD code \texttt{GORGON} \citep{Chittenden2004} is used to simulate the 2D MagLIF geometry.  A single temperature, no radiation transport, and no thermonuclear burn is assumed.  The focus of the MLDL is on reproducing and analyzing the emergent behavior of the liner dynamics.  An ensemble of 539 simulations is done, with samples at different Liner Aspect Ratios (AR) and different preheat temperatures ($T_0$), so that the compression is done on different adiabats.  Also, the ensemble has random liner perturbations in both amplitude and phase.  The evolution was sampled every 2.5 ns over 200 ns, generating 87,318 images of liner density and magnetic field.

Based on this ensemble a MLDL workflow is developed to form a surrogate and to gain insight into the nonlinear physics.  A Principle Components Analysis (PCA) is done on the logarithm of the MST of the logarithm of the liner density and the magnetic field.  That is, the logarithm of the liner density is taken, and the MST is calculated. The logarithm of the MST is then calculated, and the resulting values are subjected to a PCA. It is found that most of the variation is in the first seven components.  Singular Value Decompositions (SVDs) are done of the cross-correlations of the input parameters (AR, $T_0$, and liner perturbations) to the logarithm of the MST, and of the SVD vector components to the PCA vector components.  This allows the PCA vectors to be identified with the input parameters.  A Multi Layer Perceptron Neural Network (MLP/NN) \citep{haykin1994neural} with a three-layer encoder/decoder structure is trained to predict the seven PCA vector components as a function of time, given the initial conditions.  Excellent performance is found, as shown in Sec.~\ref{results}.  A permutation importance analysis is done on the inputs.  It is determined that the quadrupole moment quickly decays and the energy inverse cascades into the dipole moment, where it remains through stagnation and the subsequent expansion.  The evolution shows little to no dependence on the initial tripole or quadrupole moments but very strong dependence on time, AR, and the initial phase of the dipole moment, and modest dependence on the initial temperature and the size of the initial perturbation.

The analysis is repeated using the WPH in place of the MST.  While the WPH cannot be interpreted physically, as it is currently implemented, it can be inverted due to the additional phase information.  When this is done, the temporal evolution of the two fields is well-predicted, including the field-to-field correlations.

The MST and the WPH will be described in Sec.~\ref{mst.theory}, along with the display and interpretation of the MST.  How the training and testing dataset was constructed and generated appears in Sec.~\ref{dataset.generation}, and a description of the evolution is given.  The details of the MLDL architecture is given in Sec.~\ref{mldl.architecture}, followed by the results from applying the MLDL architecture in Sec.~\ref{results}.  A discussion of the results, conclusions that can be drawn, upcoming work, and possible improvements to the MLDL architecture is found in Sec.~\ref{conclusions.discussion}.

\section{Mallat Scattering Transformation (MST) and Wavelet Phase Harmonics (WPH)}
\label{mst.theory}

The Mallat Scattering Transformation (MST) can be viewed as a Convolutional Neural Network (CNN) \citep{LeCun1989,Ning2005,Goodfellow2014} with predetermined weights.  The filters are cleverly designed so that by a convolve-binate cycle, the CNN can span an exponentially large range in scale with a kernel of constant size.  In other words, a very fast algorithm, analogous to a Fast Fourier Transform (FFT), can be constructed.  For instance, on modern GPUs, the MST takes about 10 ms on a 512x512 image.  A very useful way of defining the $m$-th order MST of a field $f(x)$ is
\begin{equation}
\label{mst.eqn}
    S_m[f(x)](p) \equiv \phi_{p_\text{min}} \star \left( \prod_{k=1}^m{\text{mod} \, \psi_{p_k} \star} \right) \, f(x),
\end{equation}
where $\phi(x)$ is the Father Wavelet, $\psi(x)$ is the Mother Wavelet, mod is the complex modulus,
\begin{equation}
    \phi_p = p^2 \, \phi(px),
\end{equation}
\begin{equation}
    \psi_p = p^2 \, \psi(px),
\end{equation}
\begin{equation}
\label{wigner.weyl}
    \psi_p \star f \equiv \int{\psi_p(x') \, f(x'-x) \, dx'},
\end{equation}
\begin{equation}
    p_k \in \{2^{-j}\},
\end{equation}
\begin{equation}
    j \in \{1,2,\dots,J\},
\end{equation}
\begin{equation}
\label{normal.ordering}
    p_{k+1}<p_k,
\end{equation}
and
\begin{equation}
    p = \sum_{k=1}^\infty{p_k}.
\end{equation}
Note that this is a path integral in $p_k$ and a prescription needs to be chosen of how to go around the poles, see~\citet{zinn2010path} for additional details.  Equation~\eqref{normal.ordering} is the common choice of normal ordering.  Also note that Eq.~\eqref{wigner.weyl} is a Wigner-Weyl-like mapping from a cotangent bundle $T^*M$ with coordinates $(\pi,f)$ and symplectic metric $d\pi \wedge df$ (where $f$ is the field and $\pi$ is the canonically conjugate momentum) to the space of analytic functions on $\mathbb{C}$ with coordinate $z$.

This transform has been shown by \citet{Mallat2012} to be Lipschitz continuous to diffeomorphic deformation (unlike the Fourier Transformation), and by construction (through the final convolution with $\phi_{p_\text{min}}$) to be translationally invariant. That is, it is stationary.  The latter we view as an unfortunate step, because most physical processes are not stationary.  This is not necessary and will be discussed at length in Sec.~\ref{conclusions.discussion}.  The MST is rarely normalized in implementations despite the normalization being presented in \citet{Mallat2012} section on Dirac normalization of the MST.

The form of the MST in Eq.~\eqref{mst.eqn} can be identified as a CNN with a specified (fitting to data not necessary) and elegant structure.  There is a multi layer (in $m$, the order of the MST) application of a bank of convolutional kernels (that is, $\psi_p$), a nonlinear activation (that is, $\text{mod}$), and a final pooling operation (that is, $\phi_{p_{\text{min}}})$.  We will then do a dimensional reduction using a PCA.

\begin{figure}
\includegraphics[width=\columnwidth]{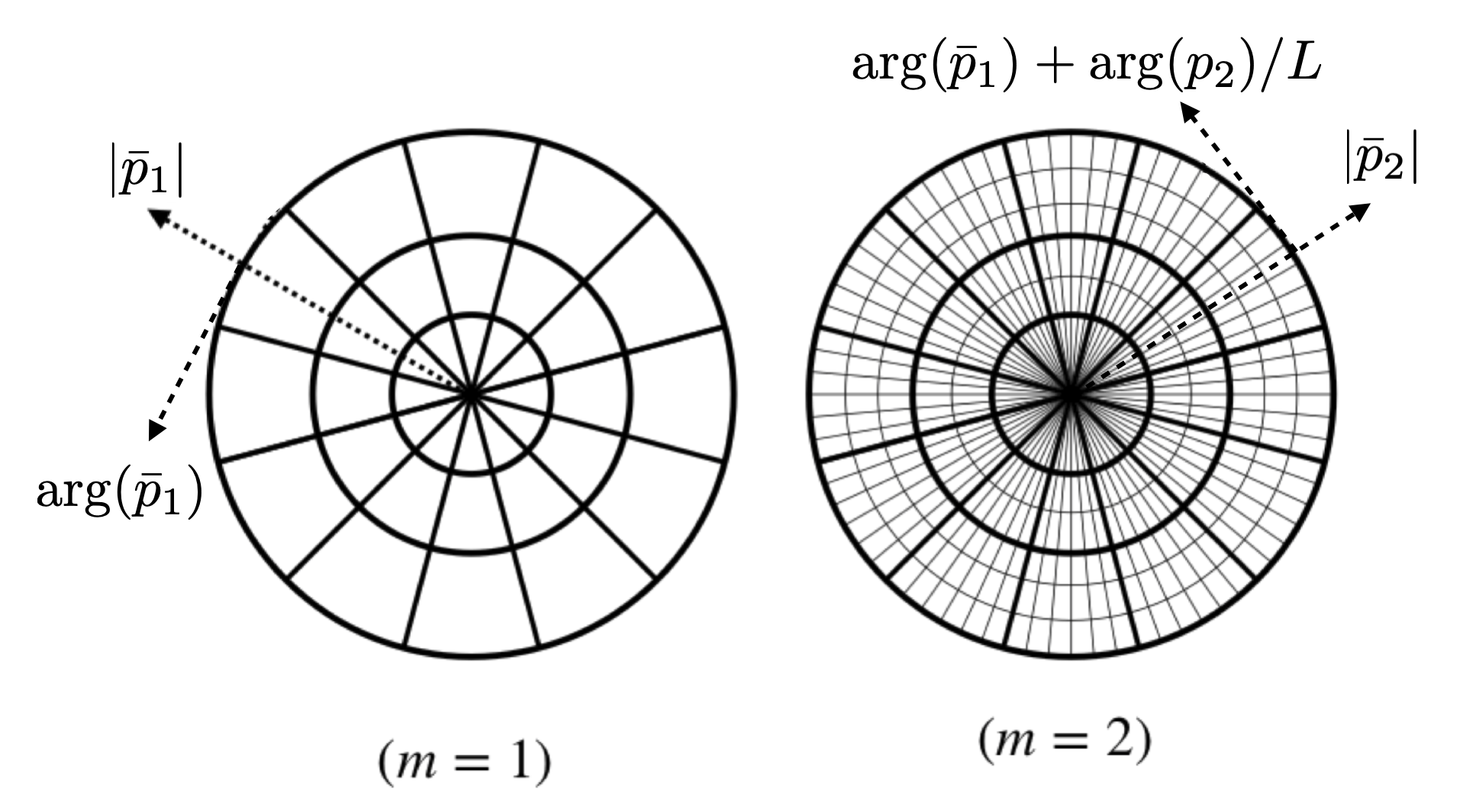}
\caption{\label{rose.plot} Coefficients produced by applying MST to 2D images in this work will be displayed on radial plots as shown. Bins are created according to scale (radial positioning, $|\bar{p}_m|$ as defined in Eq.~\eqref{bar.p}, and rotations $(\arg(\bar{p}_1),\arg(\bar{p}_1)+\arg(p_2)/L)$ for first and second order, with magnitude (color scale, not shown) representing the size of the coefficient at that scale and rotation.}
\end{figure}

\citet{Bruna2013} devised a way of visualizing the 2D MST that is shown in Fig.~\ref{rose.plot}.  The coefficients of the MST are plotted on radial plots, one for each order $m$, with the radial position being one of $|\bar{p}_m|$, $\ln|\bar{p}_m|$, $|1/\bar{p}_m|$, or $\ln(1/|\bar{p}_m|)$, where
\begin{equation}
\label{bar.p}
    \bar{p}_m \equiv \sum_{k=1}^m{p_k}.
\end{equation}
The ``posting" or manifestation of the radial plot for the angular position is much more convoluted. For $m=1$, it is simple: the angle is $\arg(\bar{p}_1)$.  Because the basis used for the 2D transform is not orthogonal, the angle for $m=2$ is calculated as $\arg(\bar{p}_1)+\arg(p_2)/L$, where $L$ is the number of angular sectors calculated.  This has the undesirable property that as the angular resolution is increased, the display is not simply a higher resolution version.
%===============================%
%===============================%

One deficiency of the MST is that it discards the phase when it takes the modulus.  This does not matter when it is used for image classification or regression, but it is a serious problem when the MST, predicted by the regression, needs to be inverted to get the predicted image.  It is analogous to inverting a Fourier transform with only the modulus.  To address this situation, Wavelet Phase Harmonics (WPH) were developed by \citet{mallat2020phase} and \citet{zhang2021maximum}.  This theory has conceptually replaced the modulus with a phase harmonic expansion
\begin{equation}
    \text{mod}(z) \xrightarrow{} \sum_{k=0}^\infty{\text{mod}(z) \, \text{e}^{\text{i} \, k \, \arg{(z)}}}.
\end{equation}
Unfortunately, there is no obvious way to plot the transformation, as it is not orthogonal and is significantly over-determined.  Because of this, a fast inverse cannot be constructed and a conventional gradient descent optimization must be done, where the objective is the $L_{2}$ norm of the difference in the WPHs.  It is extremely slow.  Where the forward WPH transform takes less than a second, the inverse transform may take more than an hour.  Even with these deficiencies, this transformation has been used with great success, most notably to analyze cosmological simulations  \citep{allys2020new,regaldo23}.

\section{Generation of the dataset}
\label{dataset.generation}

\begin{figure}
\noindent\includegraphics[width=\columnwidth]{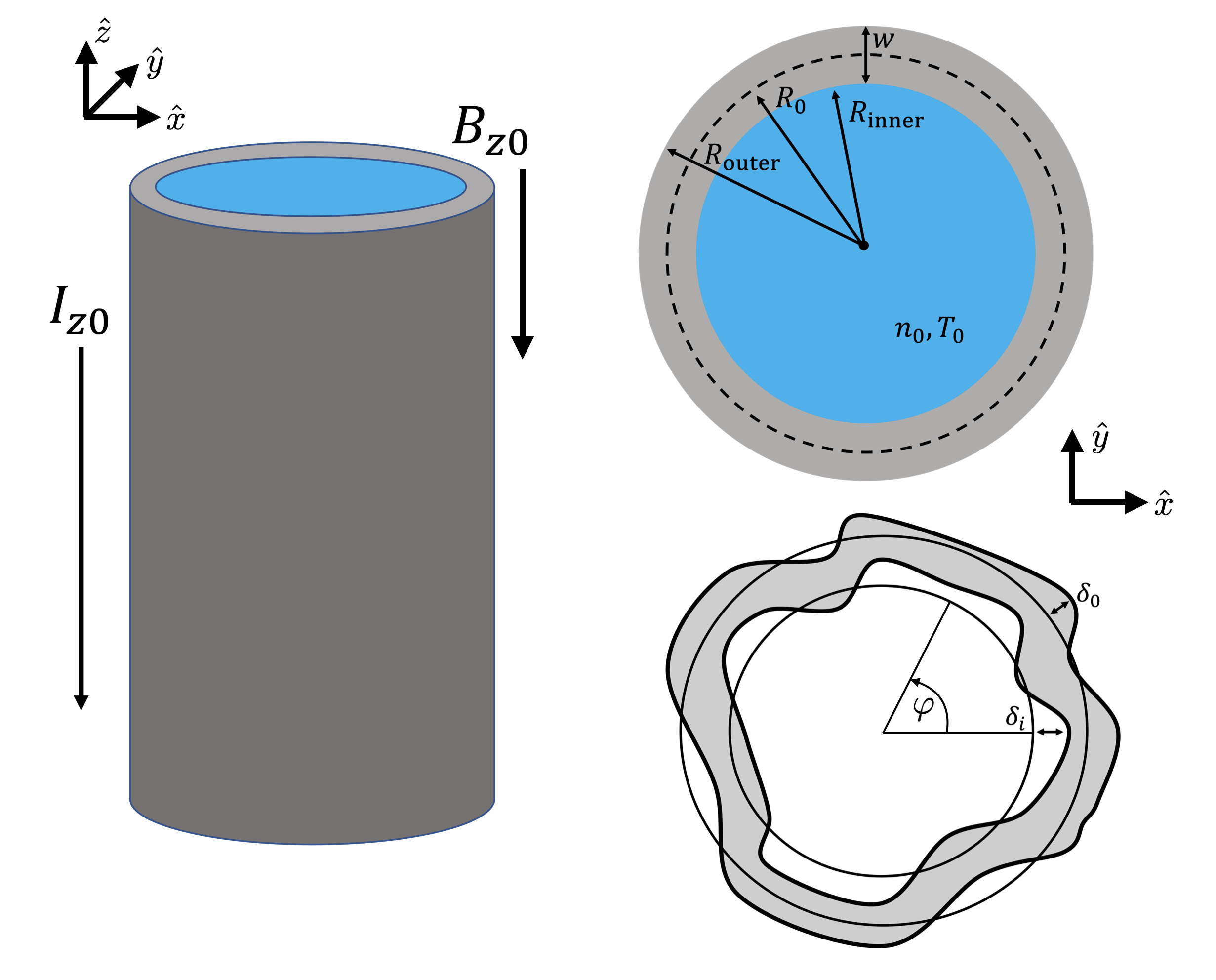}
\caption{\label{simulation.geometry} Geometry of the simulations.}
\end{figure}

An ensemble of 2D resistive MHD simulations were done using the finite volume \texttt{GORGON} code.  Axial symmetry was assumed and the simulation was done in the 2D $(x,y)$-plane as shown in Fig.~\ref{simulation.geometry}.  A geometry relevant to MagLIF was used with a cylindrical beryllium liner with inner and outer radii 
\begin{equation}
\begin{split}
    R_{\text{outer}} &\equiv \left( 1+\frac{w}{2 \, R_0} \right) \, R_0, \\
    R_{\text{inner}} &\equiv \left( 1-\frac{w}{2 \, R_0} \right) \, R_0,
\end{split}
\end{equation}
where $R_0$ is the mean radius and $w$ is the liner thickness.  We define the liner aspect ratio as $\text{AR}= R_0/w$.  Note that the larger the AR, the larger the liner acceleration, and, therefore, the larger the linear growth rate of the MRT.  The liner inner and outer surface is perturbed with dipole (m=2) to quadruple (m=4) azimuthal perturbations to seed the MRT of form
\begin{equation}
\begin{split}
    \delta_o &= \sum_{m=2}^{4}{\delta_{o0m} \cos(2 \, \pi \, m \, \varphi \, + \, \varphi_{om})}, \\
    \delta_i &= \sum_{m=2}^{4}{\delta_{i0m} \cos(2 \, \pi \, m \, \varphi \, + \, \varphi_{im})}.
\end{split}
\end{equation}
Inside the liner is a Deuterium ($D_2$) gas with initial density $n_0$, preheated to a temperature $T_0$.  Note that a larger value of $T_0$ will put the implosion on a lower adiabat, causing the implosion to reach lesser compression ratios and a larger radius at stagnation.  This gives the MRT less time to grow.  A uniform axial magnetic field $B_{z0}$ is initially established, then a large axial current with a sinusoidal profile is driven through the liner, with a peak current of $I_{z0}$ at $100 \, \text{ns}$ and a total duration of $200 \, \text{ns}$.
%===============================%
%===============================%

An ensemble of 539 simulations were done with the liner density $n_l(x,y;t)$ and magnetic field strength $B(x,y;t)$ sampled every $2.5 \, \text{ns}$ with a $10 \, \mu m$ grid spacing in $x$ and $y$.  This took over 200k core$\ast$hrs on a high performance cluster at LLNL, and generated $87,318$ 256$\times$256 images.  To reduce the number of parameters and simplify the analysis, the liner perturbations were limited to $w \,\Delta = \delta_{i0m} = - \delta_{o0m}$ and $\varphi_{im}=\varphi_{om}$ for $m=2,3,4$.  The parameters $R_0=2.4 \, \text{mm}$, $B_{z0}=10 \, \text{T}$, $n_0=1 \, \text{mg/cc}$, and $I_{z0}=20 \, \text{MA}$ were held constant.  That left 6 parameters to be sampled, including the initial conditions and the stochastic parameters.  The two initial conditions included $\text{AR}=[3,9]$ and $\log_{10} \, T_0 = [1,2.8]$ (that is, $T_0 = [10 \,\text{eV}, 630 \, \text{eV}]$). The four stochastic parameters included $\log_{10} \, \Delta =[-2,-1]$ (that is, $\Delta = [1  \%,10   \%]$) and $\varphi_{im}=[0,2 \pi]$ for each of $m=2,3,4$.  The smallest $\Delta$ corresponds to $1\%$ of the thinnest liner, which is about $5 \, \mu m$, about half a grid cell.  Latin Hypercube Sampling (LHS) was used to generate 27 samples of $(\text{AR}, T_0, \Delta)$, with the $\varphi_{im}$ being chosen from uniform distributions.  Another 512 parameter vectors were randomly sampled from their uniform distributions.

\begin{figure}
\noindent\includegraphics[width=\columnwidth]{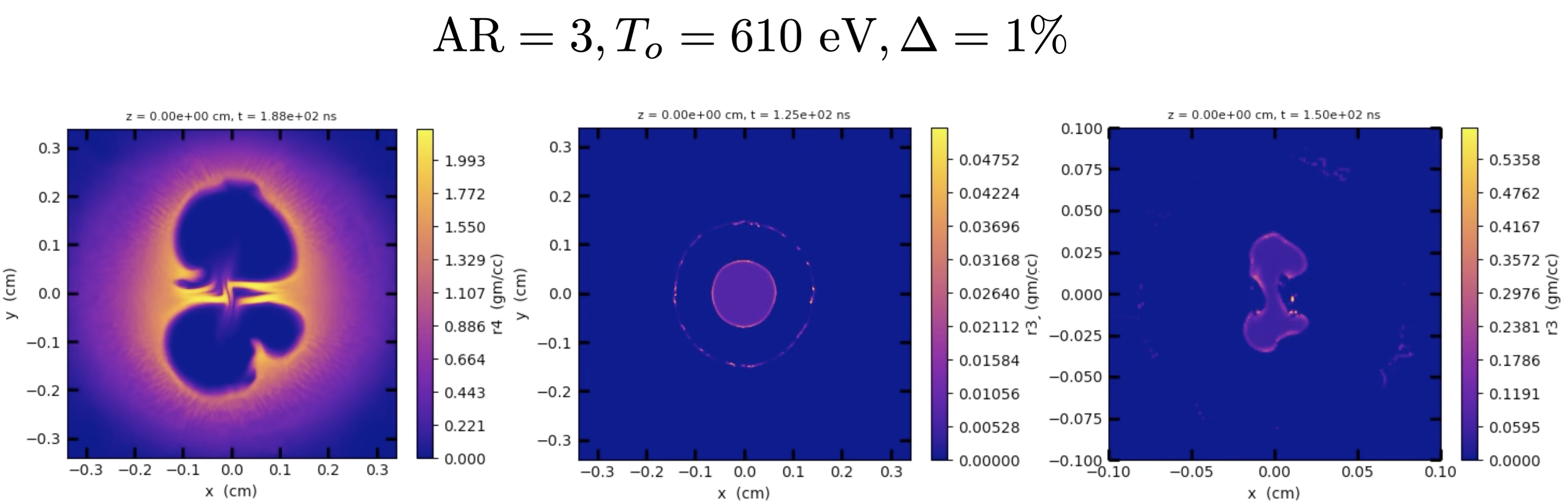}
\caption{\label{gorgon.evolution} Evolution of the liner density $n_l(x,y;t)$ and the gas density $n_g(x,y;t)$ as simulated by $\texttt{GORGON}$.  The picture on the right is a zoom in on the stagnation of the gas.  The animation of this figure can be found at this link to a \href{https://youtu.be/GmIr3O5GLR0}{Multimedia View}.}
\end{figure}

An example of one of the evolutions of the liner density $n_l$ and the gas density $n_g$ is shown in Fig.~\ref{gorgon.evolution}.  This simulation was done with an $\text{AR}=3$, $T_0=631 \, \text{eV}$, and $\Delta=1\%$.  Note how gas stagnates into a dipole pattern in a sausage-like mode, then that pattern is imprinted on the liner as it expands post stagnation.
%===============================%
%===============================%

\section{MLDL architecture}
\label{mldl.architecture}

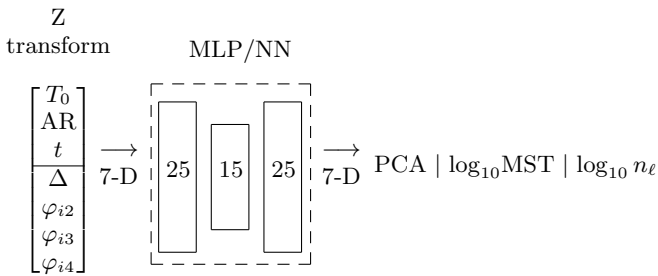
\begin{figure}
\vspace{1cm}
%\noindent\includegraphics[width=\columnwidth]{figures/mldl_workflow.png}
%\hspace{-0.5cm}
\hspace{-0.15cm}
\begin{minipage}{0.08\textwidth}
\vspace{-0.6cm}
\begin{center}
Z transform
\end{center}
\[
\displaystyle
\begin{bmatrix}
T_{0}       \\
\text{AR}          \\
t           \\ \hline
\Delta      \\
\varphi_{i2}   \\
\varphi_{i3}   \\
\varphi_{i4}
\end{bmatrix}
\]
\end{minipage}
\hspace{-0.4cm}
\begin{minipage}{0.03\textwidth}
$\longrightarrow$ \\ 7-D
\end{minipage}
\hspace{-0.6cm}
\begin{minipage}{0.15\textwidth}
\setlength{\unitlength}{0.2cm}
\begin{picture}(3,13.5)
    \put(0,14){MLP/NN}
    \multiput(-2.5,0.2)(0,1.4){9}{\line(0,1){0.8}}
    \multiput(-2.5,0.2)(1.4,0){8}{\line(1,0){0.8}}
    \multiput(-2.5,12.2)(1.4,0){8}{\line(1,0){0.8}}
    \multiput(8.1,0.2)(0,1.4){9}{\line(0,1){0.8}}
    \put(-1.5, 6){25}
    \put(-2, 1){\line(0,1){10}}
    \put(-2, 1){\line(1,0){2.5}}
    \put(0.5, 1){\line(0,1){10}}
    \put(-2, 11){\line(1,0){2.5}}
    \put(2, 6){15}
    \put(1.5, 2.5){\line(0,1){7}}
    \put(1.5, 2.5){\line(1,0){2.5}}
    \put(4, 2.5){\line(0,1){7}}
    \put(1.5, 9.5){\line(1,0){2.5}}
    \put(5.5, 6){25}
    \put(5, 1){\line(0,1){10}}
    \put(5, 1){\line(1,0){2.5}}
    \put(7.5, 1){\line(0,1){10}}
    \put(5, 11){\line(1,0){2.5}}
\end{picture}
\end{minipage}
\hspace{-0.1cm}
\begin{minipage}{0.03\textwidth}
$\longrightarrow$ \\ 7-D
\end{minipage}
\hspace{-0.05cm}
\begin{minipage}{0.21\textwidth}
PCA $\mid$ $\log_{10}$MST $\mid$ $\log_{10}n_{\ell}$
\end{minipage}
\caption{\label{mldl.workflow} The MLDL workflow.}
\end{figure}

The following MLDL workflow was constructed and used to analyze the data.  The logic behind the construction will be discussed in Sec.~\ref{conclusions.discussion}.  The pipeline is shown in Fig.~\ref{mldl.workflow}.  First, the logarithm of each pixel in the images was calculated, and the MST and WPH were subsequently found for the logarithmic image. It took 16 GPU$\ast$hrs (on Nvidia GTX 3090) to take the MSTs of the images and 27 GPU$\ast$hrs to take the WPH of the images.  Values of $J=8$ and $L=16$ were used to take the MST, and values of $J=L=8$, $\Delta_j=5$, $\Delta_k=0$, $\Delta_l=8$, and $\Delta_n=2$ to take the WPH.  Though these key specifications are not discussed in this paper, they are included here for completeness and reproducibility; details can be found in~\citet{Mallat2012} and~\citet{mallat2020phase} for MST and WPH, respectively. The logarithm (base 10) was then taken of the MST and the mean was subtracted on a coefficient by coefficient basis.  Note that subtracting the mean from the $\log_{10}$ is equivalent to applying a multiplicative scaling to the original transform.  It is known that this implementation of the MST is not properly Dirac-normalized.  This leads to an imprinted pattern that distracts from the natural variation in the image.  Subtacting the mean removes this imprinted pattern.  It should be noted that this does not affect the subsequent PCA.  The analysis of the MST is followed by a PCA.  The data is projected onto the top seven PCA vectors for subsequent analysis.  On the other side of the pipeline, the input vector of the three control variables $(\text{AR},T_0,t)$ and the four stochastic variables $(\Delta,\varphi_{i2},\varphi_{i3}.\varphi_{i4})$ are Z-normalized.
%===============================%
%===============================%

In order to identify the PCA vectors, an SVD is done on the cross-correlation of the inputs to the output, before the projection onto the PCA vectors.  This allows for a labeling of the SVD vectors according to the inputs.  A second SVD is taken of the cross-correlation between the projection onto the PCA vectors and the projection onto the SVD vectors.  This allows the labeled SVD vectors to be associated with the correct PCA vector, thereby labeling the PCA vectors with the input.

The last step is to train a MLP neural network with ReLU activation to predict the seven PCA vector components given the inputs.  The structure of the MLP/NN was optimized, as well as the regularization parameter, via a grid search.  There was a bias in this choice toward more regularization (without significantly compromising performance) to prevent over-training.  The optimal structure was one with three hidden layers (with 25-15-25 nodes) with a encoder/decoder structure.  A permutation importance analysis was also done to determine the importance of the input parameters in the estimation.  A five-fold cross validation was used in all analyses.  As will be shown in the next section, the MLP/NN had very good performance, and a very interpretable result.  A shallow Support Vector Regression (SVR) with a radial basis was attempted.  The results were very low frequency and were unable to capture the details of the stagnation.  A higher frequency result could be obtained by decreasing the regularization, but the result did not cross-validate.

For this application, the primary downside to the MST is that it throws out the phase of the transformation.  While this does not matter for classification or some of the physical interpretation, it prevents the transform from having a useful inversion.  It is akin to throwing out the phase of a Fourier Transform, then inverting with a random phase.  For this reason we repeated our analysis using the WPH (the MST with phase).  While this transformation does have reasonable inversions, it has no physically interpretable display.  It is a black-box vector.  Care also needed to be taken with the treatment of the complex-valued transform.  The complex analytic $\ln(z)$ function was used, yielding $\ln |z| + \text{i} \, \arg(z)$, and a circular correlation \citep{jamm2001} was used in the PCA with respect to the cyclic $\text{Im}(\ln(z))$.  Finally, due to the unreasonable translational invariance built into the WPH, there is an arbitrary $x$ and $y$ translation that must be removed. A fiducial at the vertical and horizontal edges of the images was added to aid in this task.

The PCA only took 1 core$\ast$sec, and the MLP/NN took 20 core$\ast$sec to train.  The resulting forward model surrogate takes 0.1 core$\ast$sec to evaluate, compared to the 360 core$\ast$hrs required for the $\texttt{GORGON}$ simulation -- a factor of $10^7$ acceleration.

The implementation of the MST used was the \texttt{Kymatio} package \citep{andreux2020} which available on \href{https://github.com/kymatio/kymatio}{Github}. The version of the WPH used is based on the work of \citet{zhang2021maximum} which is also available on \href{https://github.com/bregaldo/pywph}{Github} as the \texttt{pyWPH} package.

\section{Results}
\label{results}

\begin{figure}
\noindent\includegraphics[width=\columnwidth]{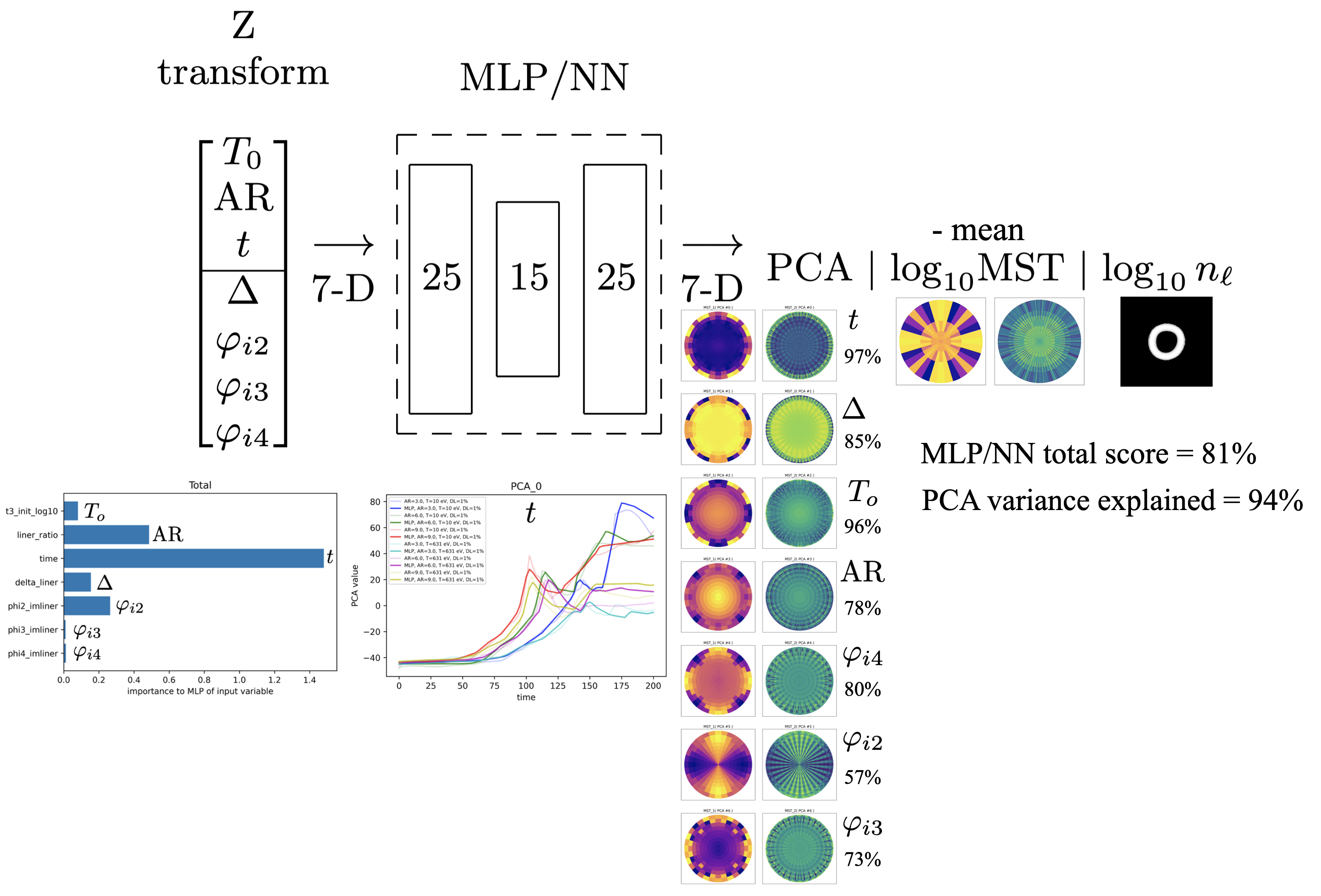}
\caption{\label{mldl.workflow.total} The complete MLDL workflow with key results.  Underneath a diagram of the MLDL pipeline, going right to left, are: (1) grey scale image of $n_l$, (2) first and second order MST of the grey scale image, (3) the MST PCA vectors, (4) some predictions of the first PCA vector component with respect to time, and (5) the permutation feature importance analysis.}
\end{figure}

An overview of the results are shown in Fig.~\ref{mldl.workflow.total}.  Images corresponding to the MLDL pipeline are shown beneath the schematic of Fig.~\ref{mldl.workflow}.  On the right-hand side is the $\log_{10}$ image of the liner density.  To the left of it are the first and second order MST of that image.  Continuing to the left of that are the seven principal vectors of the MST, which explain $94\%$ of the variance.  The identification of the PCA vectors with the inputs along with the five-fold cross validated score for each PCA vector, as predicted by the MLP/NN, are included.  The total score was $81\%$.  A typical example of the PCA vector evolution with respect to time is shown, as predicted by the MLP/NN.  Finally, the permutation feature importance result is shown on the left of the image.  More details of these results follow, and they will be discussed in detail in the following section.
%===============================%
%===============================%

\begin{figure}
\noindent\includegraphics[width=\columnwidth]{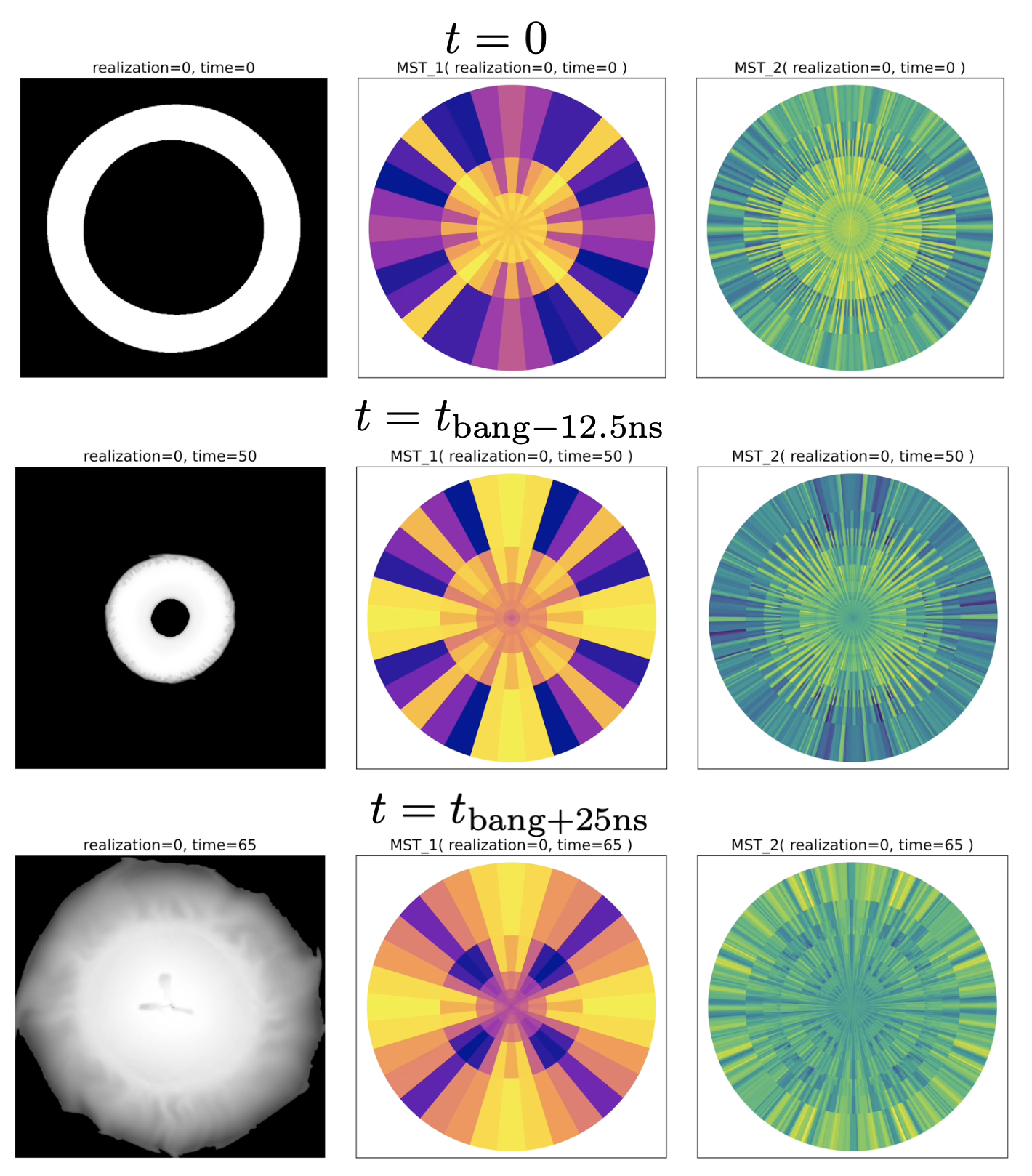}
\caption{\label{mst.times} Ensemble of realizations at three select times.  Shown are the $\log_{10} n_l(x,y;t)$, then the first and second order MST of the liner density.  From top to bottom: (1) $t=0$, (2) $t=t_{\text{bang}-12.5 \, \text{ns}}$, and (3) $t=t_{\text{bang}+25 \, \text{ns}}$.  The radial axis is linear in scale.  The animation of this figure can be found at this link to a \href{https://youtu.be/b-p09GZigNA}{Multimedia View}.}
\end{figure}

\begin{figure}
\noindent\includegraphics[width=\columnwidth]{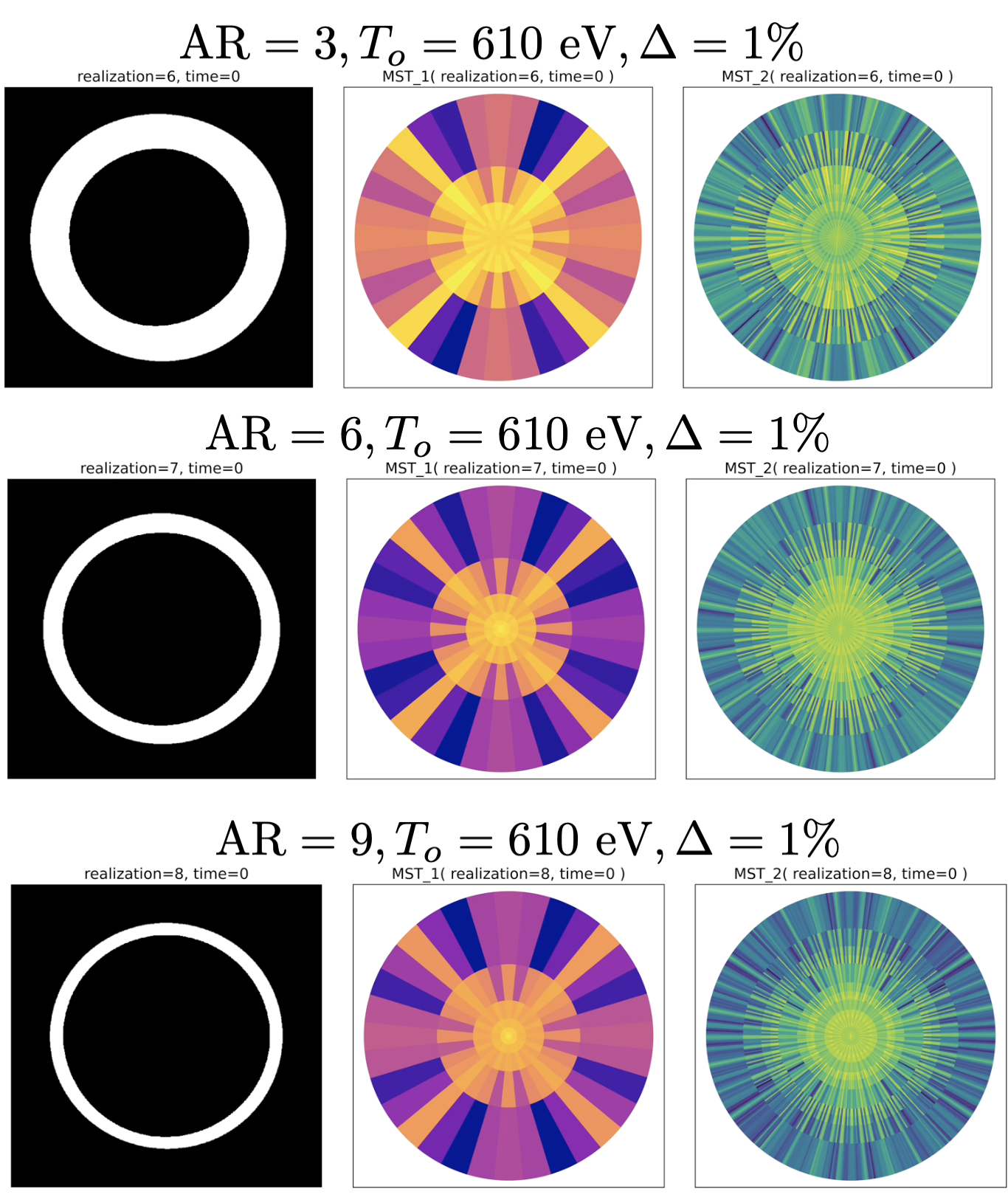}
\caption{\label{mst.evolutions} Three time evolutions of the MST of liner density.  Shown for $T_0=0$ and $\Delta=1\%$.  From top to bottom: (1) $\text{AR}=3$, (2) $\text{AR}=6$, and (3) $\text{AR}=9$.  The radial axis is linear in scale.  The animation of this figure can be found at this link to a \href{https://youtu.be/b-p09GZigNA}{Multimedia View}.}
\end{figure}

The analysis starts by taking the MST of all the images.  Shown in Fig.~\ref{mst.times} is the MST of a simulation at three different times:  at time zero, at $12.5 \, \text{ns}$ before bang time (maximum compression), and at $25 \, \text{ns}$ after bang time.  There is also a \href{https://youtu.be/b-p09GZigNA}{link} to an animation showing all of the realizations.  It is followed by Fig.~\ref{mst.evolutions}, which shows the MST at time zero of three simulations with the highest adiabat of $T_0=631 \, \text{eV}$ and the smallest liner perturbation of $\Delta=1\%$, for three different values of $\text{AR}=3,6,9$.  There is a \href{https://youtu.be/b-p09GZigNA}{link} in the figure caption to an animation of the full time evolution of each case.
%===============================%
%===============================%
%===============================%
%===============================%

\begin{figure}
\noindent\includegraphics[width=\columnwidth]{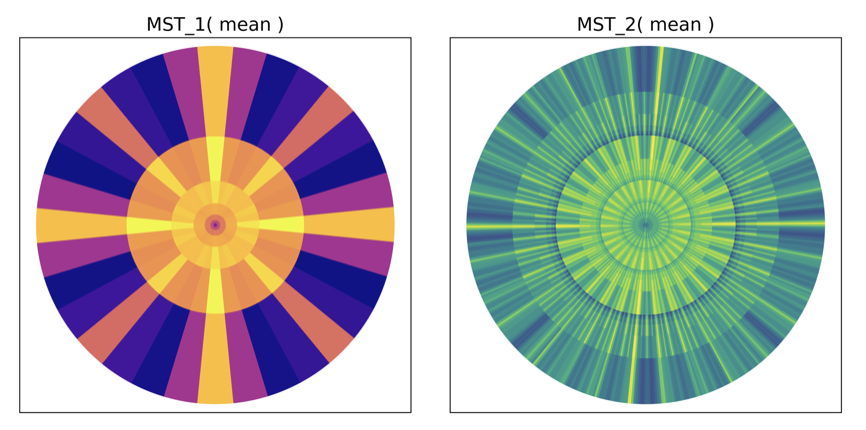}
\caption{\label{mean.mst} The mean MST of liner density.  Shown is the first order MST on the left and the second order MST on the right.  The radial axis is linear in scale.}
\end{figure}

\begin{figure}
\noindent\includegraphics[width=\columnwidth]{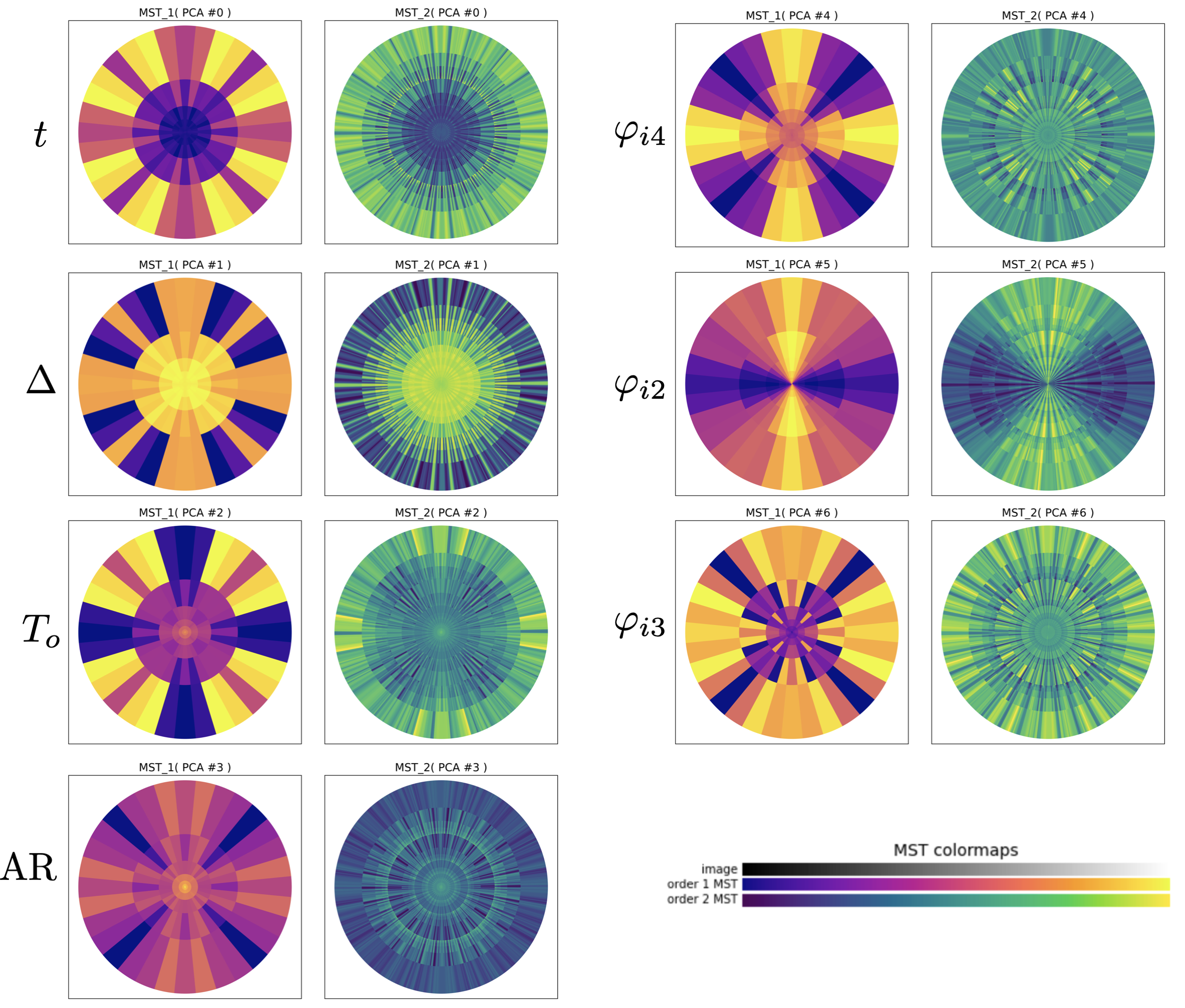}
\caption{\label{pca.vectors} PCA vectors.  Shown are the PCA vectors of liner density in MST space.  The radial axis is linear in scale.  The color bar for all MST displays is in the lower right hand corner.  The identification of each PCA vector with an input is displayed to the left of the PCA vector.  See Fig.~\ref{pca.svd.cross.analysis} for this interpretation key.}
\end{figure}

\begin{figure}
\noindent\includegraphics[width=\columnwidth]{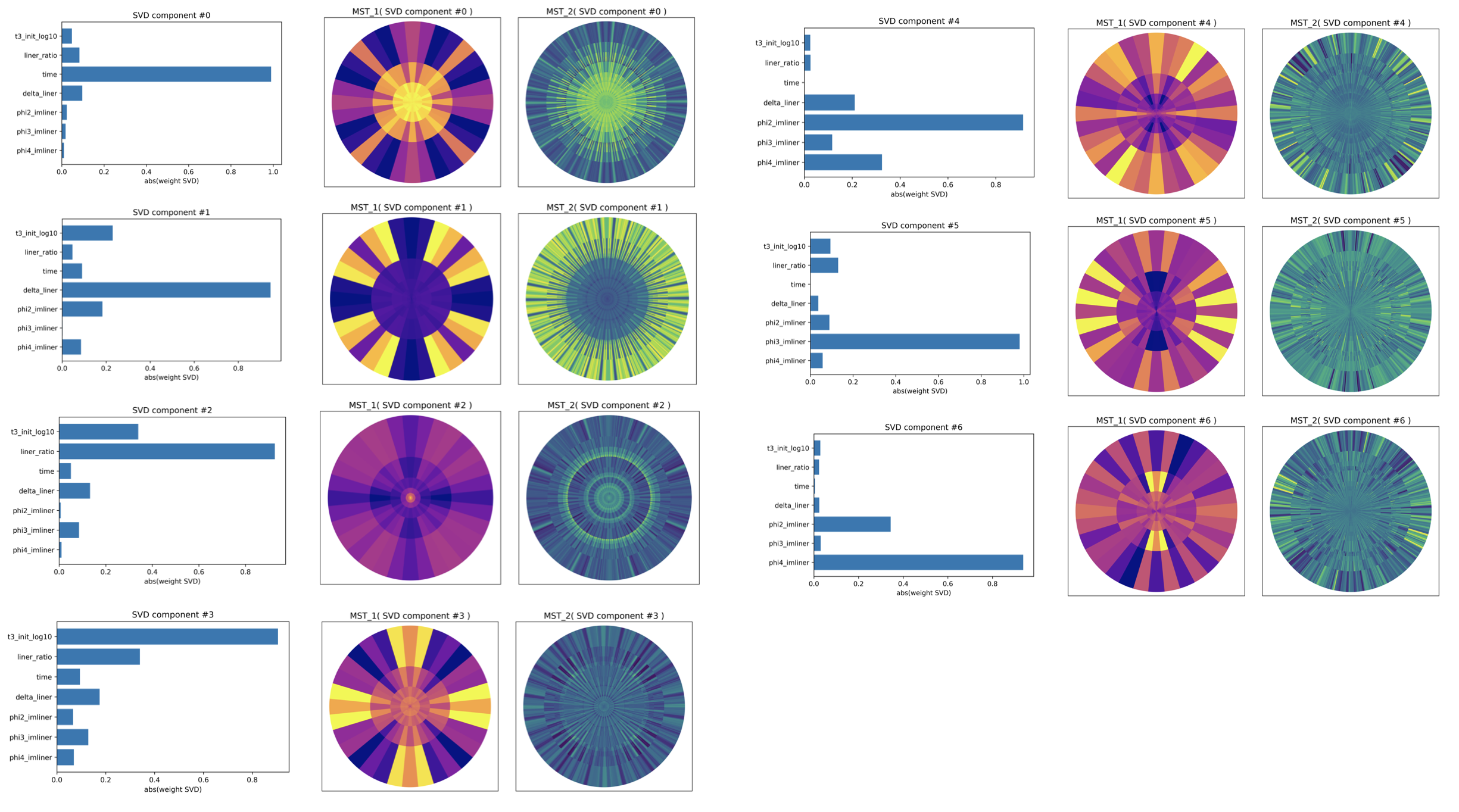}
\caption{\label{svd.vectors} SVD vectors.  Shown are the SVD vectors in both input space and liner density MST space.  Note the clear identification of each SVD component with one input.  The radial axis is linear in scale.}
\end{figure}

\begin{figure}
\noindent\includegraphics[width=\columnwidth]{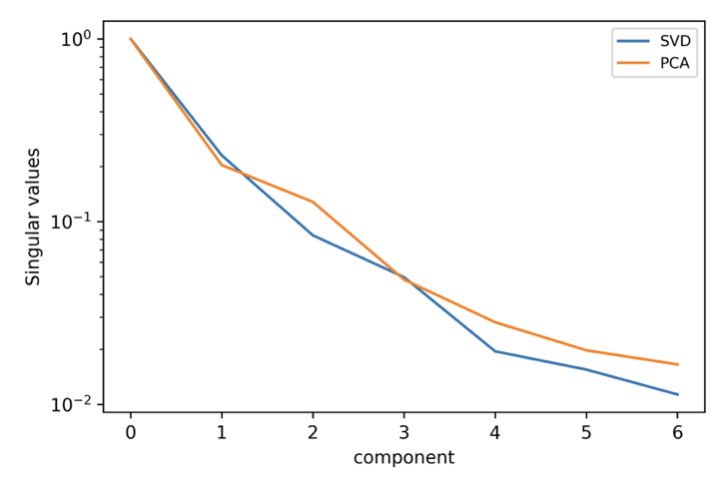}
\caption{\label{pca.svd.values} The eigenvalues of the PCA vectors compared to those of the SVD.  Note that the PCA is more efficient at capturing the variance in fewer components than the SVD.}
\end{figure}

\begin{figure}
\noindent\includegraphics[width=\columnwidth]{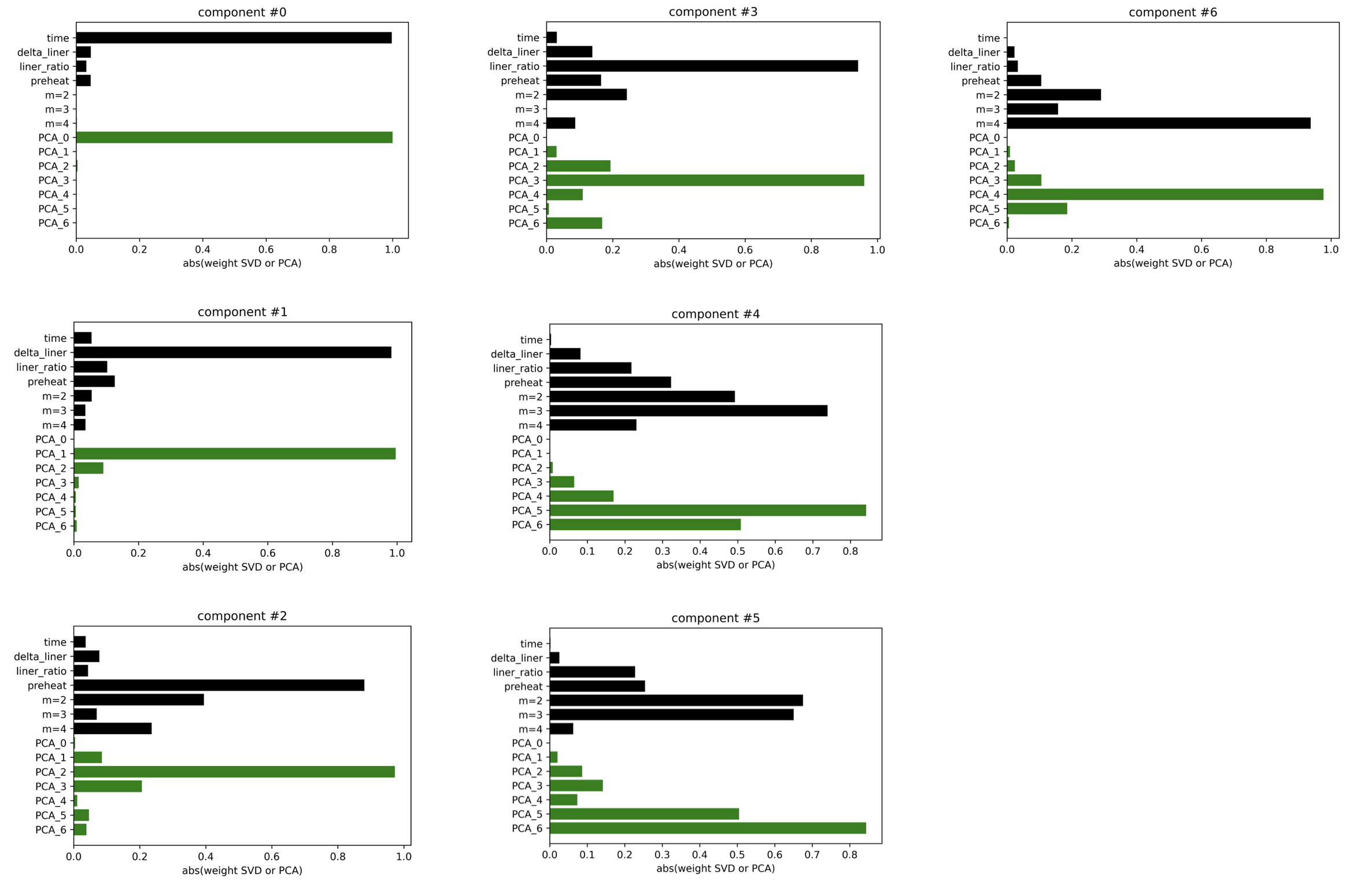}
\caption{\label{pca.svd.cross.analysis} The singular vectors of the PCA-to-SVD cross variance analysis.  This analysis allows the identification of the PCA vectors with a unique input.}
\end{figure}

The mean value for the MST (shown in Fig.~\ref{mean.mst}) is then subtracted from the values, and a PCA is done, as well as an SVD analysis of the cross correlation of the inputs to the outputs.  The PCA vectors are shown in Fig.~\ref{pca.vectors}, and the SVD vectors are shown in Fig.~\ref{svd.vectors}.  The eigenvalues for both decompositions are shown in Fig.~\ref{pca.svd.values}.  Note that the PCA is capturing more of the variation in fewer components.  The first seven PCA vectors have captured $94\%$ of the total variation.  The SVD has a strong correlation of each component with one input parameter.  When the PCA projection of the output is cross-correlated with the SVD projection of the output, an interpretation key is generated for the PCA vectors by the display of the singular vectors, as shown in Fig.~\ref{pca.svd.cross.analysis}.  %These PCA identifications are displayed on Fig.~\ref{pca.vectors}.
%===============================%
%===============================%
%===============================%
%===============================%
%===============================%
%===============================%
%===============================%
%===============================%
%===============================%
%===============================%

\begin{figure}
\noindent\includegraphics[width=\columnwidth]{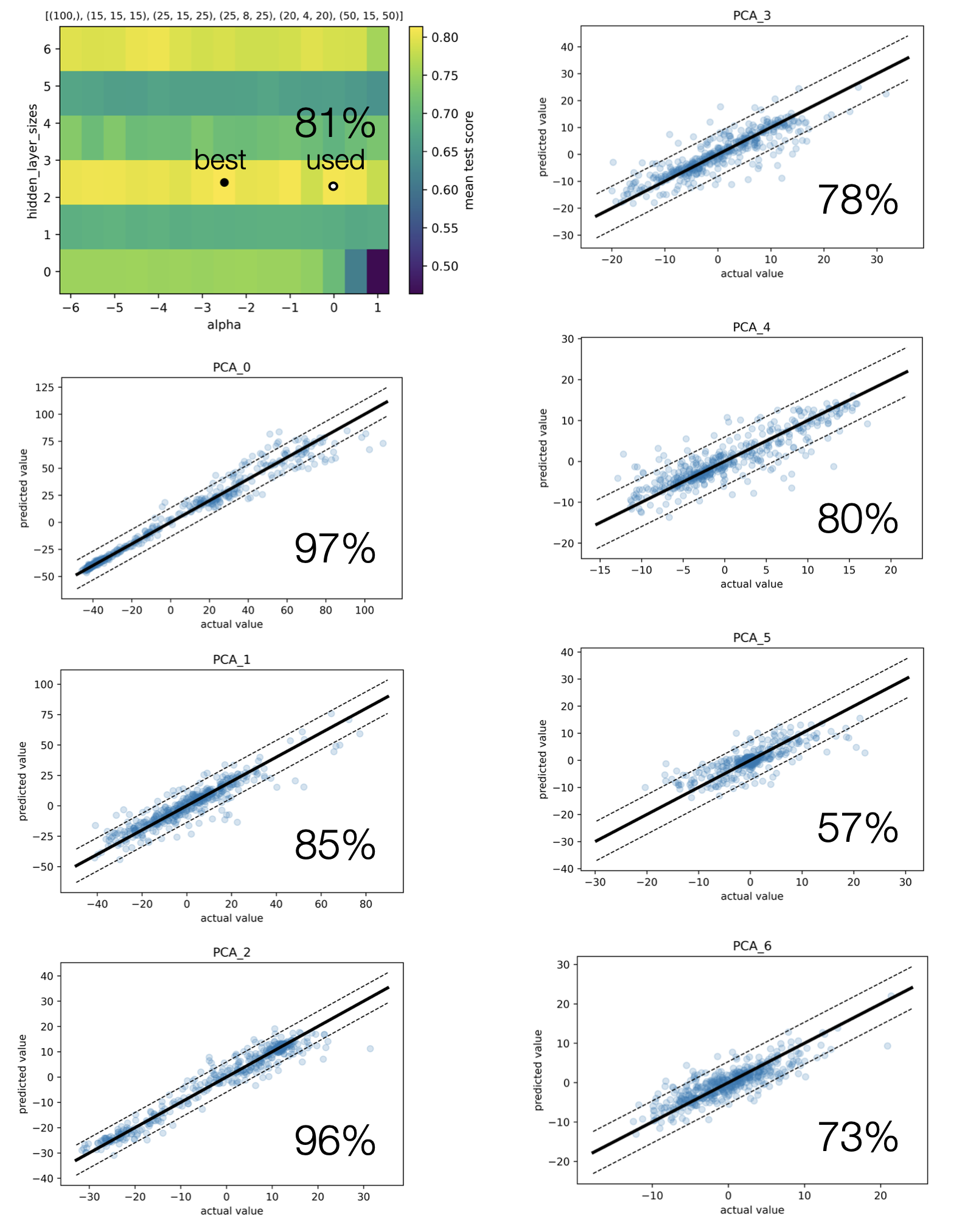}
\caption{\label{mlp.cross.plots} Cross plots of predicted versus actual values for each PCA vector  component, by the MLP/NN.  The score for each PCA vector component is shown on the respective plot.  The grid search for the MLP/NN structure and regularization parameter $\alpha$ is shown in the upper left corner.  A structure of hidden layers of  25-15-25, and a value of $\alpha=0$ is used.}
\end{figure}

\begin{figure}
\noindent\includegraphics[width=\columnwidth]{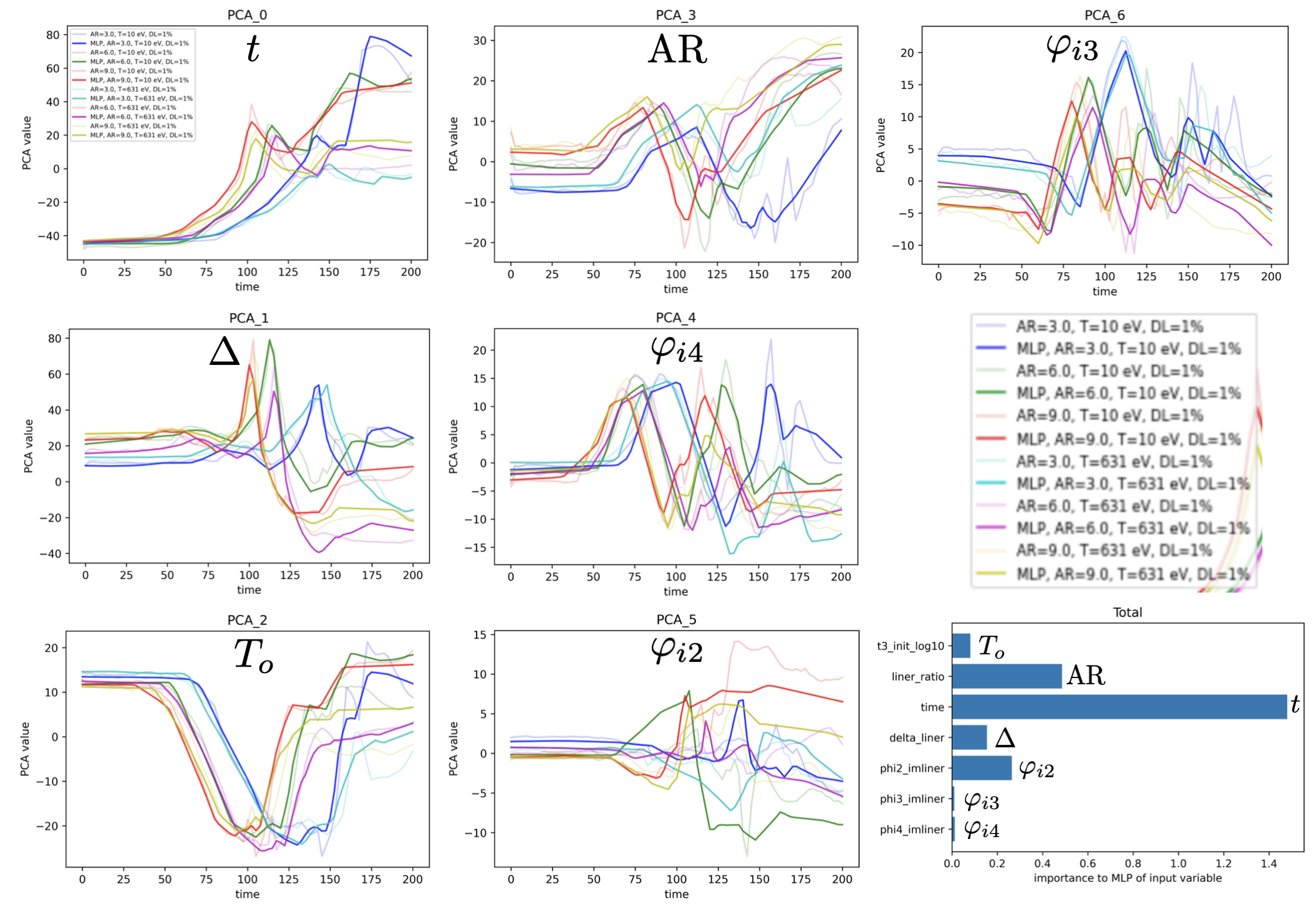}
\caption{\label{mlp.pca.evolutions} The evolution of the PCA vector components as predicted by the MLP/NN compared to the actual values.  The predicted values are the bold lines and the actual values are the light lines of the same color.  The permutation feature importance is shown in the lower right hand corner.  The identification of each PCA vector component with the unique input parameter is indicated on each plot.  Shown are six simulations that span $\text{AR}=3,6,9$ and $T_0= 10 \, \text{eV}, 631 \, \text{eV}$.}
\end{figure}

A MLP/NN is then trained to predict the PCA vector components of the fields $n_l(x,y;t)$ and $B(x,y;t)$, given the six initial condition parameters.  The performance is shown in Fig.~\ref{mlp.cross.plots}.  The results of a grid search for network structure and the regularization parameter $\alpha$ are shown in the upper left-hand corner.  Note that the value of $\alpha$ is increased from the best parameter to one with almost the same performance but more regularization in order to prevent overfitting.  The overall performance is $81\%$ and the performance on individual PCA vector components range from a high of $97\%$ to a low of $57\%$.  The performance is quantified by the linear correlation of the predicted to the actual values.  All of the correlations look very linear with no pathologies.  The performance of the MLP/NN on predicting the PCA vector components is quite remarkable and is shown in Fig.~\ref{mlp.pca.evolutions}.  Shown are the predicted versus actual for six simulations that span $\text{AR}=3,6,9$ and $T_0= 10 \, \text{eV}, 631 \, \text{eV}$.  Note that the MLP/NN decided to use few points where the function was linear and captured the stagnation behavior well where the function is singular.  The MLP/NN found the stagnation points and put an interpolation point at the cusp.  The permutation feature importance shows interesting results that will be discussed in the next section.  Do note that $\text{AR}$, $t$, and $\varphi_{i2}$ are the most important, and $\varphi_{i3}$ and $\varphi_{i4}$ are of little importance.
%===============================%
%===============================%
%===============================%
%===============================%

\begin{figure}
\noindent\includegraphics[width=\columnwidth]{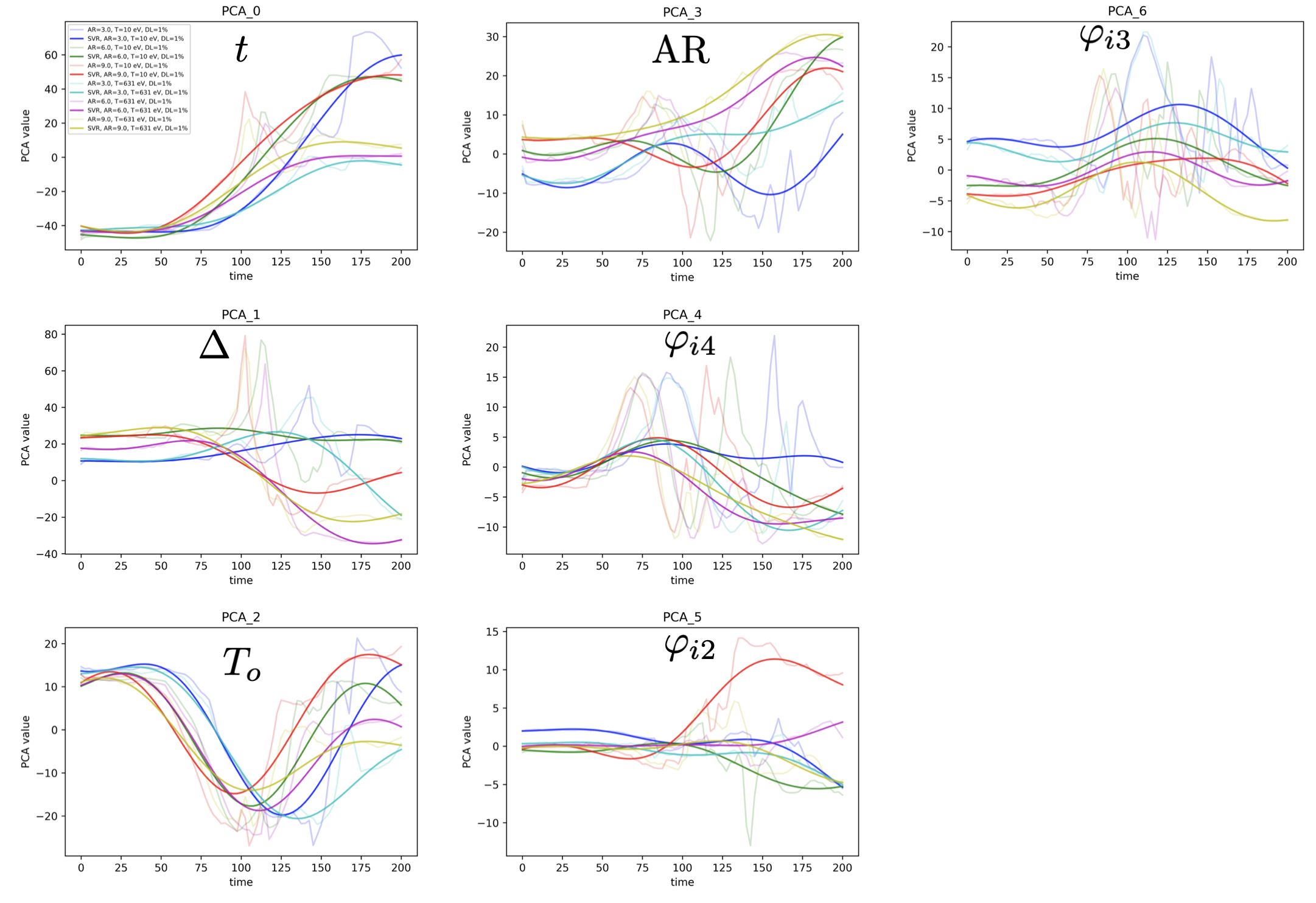}
\caption{\label{svm.pca.evolutions} The evolution of the PCA vector components as predicted by the SVM compared to the actual values.  The predicted values are the bold lines and the actual values are the light lines of the same color.  The identification of each PCA vector component with the unique input parameter is indicated on each plot.  Shown are six simulations that span $\text{AR}=3,6,9$ and $T_0= 10 \, \text{eV}, 631 \, \text{eV}$.}
\end{figure}

A Support Vector Regression (SVR) with a radial basis was attempted.  The results were disappointing.  The overall performance was $46\%$, with a high of $90\%$ and a low of $10\%$.  The predictions of the PCA vector components are shown in Fig.~\ref{svm.pca.evolutions}.  Note how (overly) smooth the regressions are, and how the stagnation (singularity) is missed completely.  When the regularization was reduced to reduce the smoothing, the performance went down significantly evidenced by the result not cross validating.
%===============================%
%===============================%

\begin{figure}
\noindent\includegraphics[width=\columnwidth]{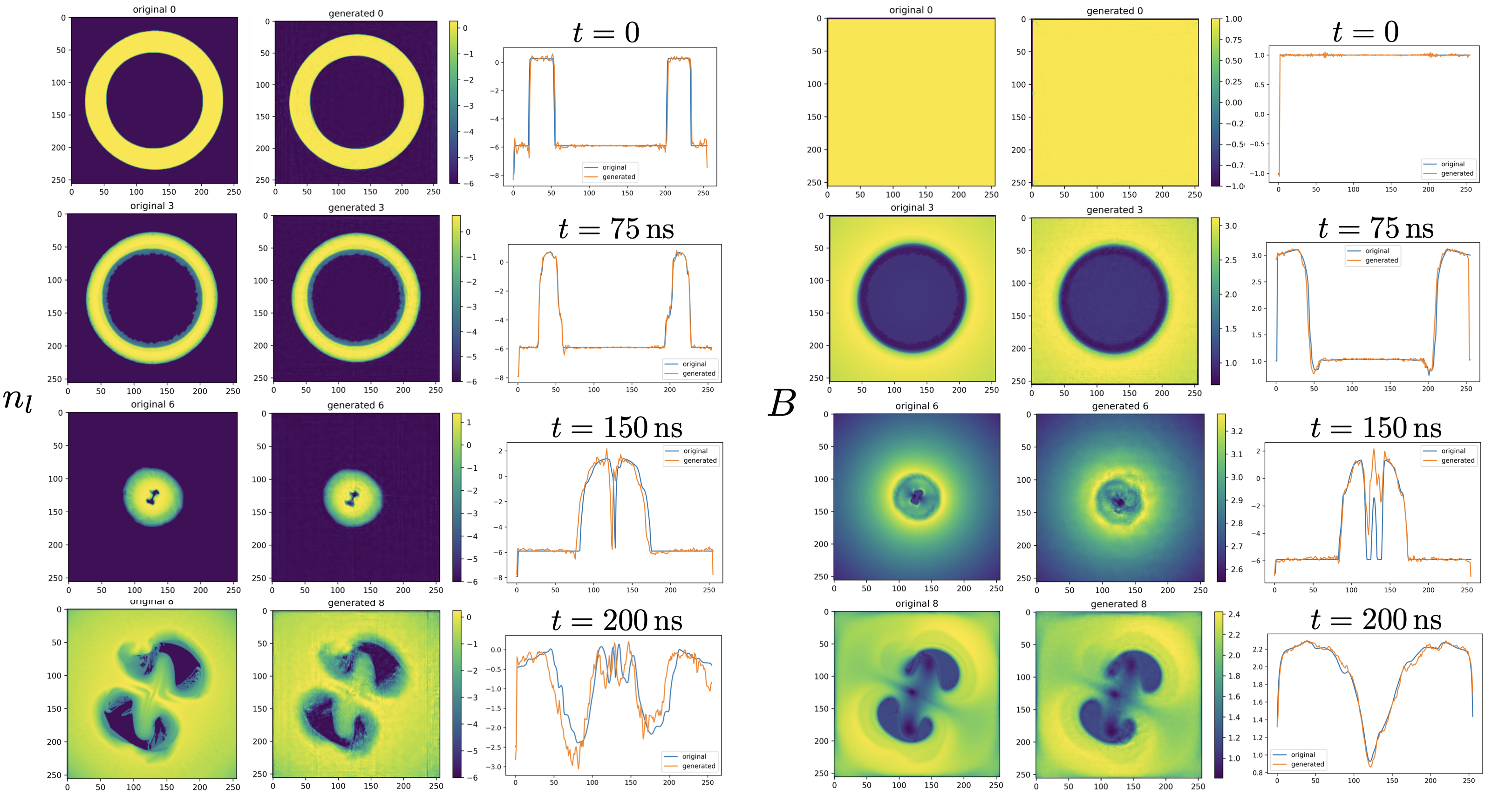}
\caption{\label{wph.results} The evolution of the liner density (left) and the magnetic field (right), and the reconstruction from the inverse WPH (iWPH) and horizontal lineouts through both profiles.  The fields are displayed at four key times (top to bottom): $t=0$, $t=75 \, \text{ns}$, $t=150 \, \text{ns}$, and $t=200 \, \text{ns}$.  The initial conditions are $\text{AR}=3$, $T_0=631 \, \text{eV}$, and $\Delta=1\%$.}
\end{figure}

Finally, the mapping back to the fields was done using the WPH.  The analysis previously described was repeated using the WPH in place of the MST.  The evolution was predicted by the MLP/NN, and the results were inverted to give the evolution at four key times for both the liner density and the magnetic field demonstrating the correlation.  The results are shown in Fig.~\ref{wph.results}.
%===============================%
%===============================%

\section{Conclusions and Discussion}
\label{conclusions.discussion}

\begin{figure}
\noindent\includegraphics[width=\columnwidth]{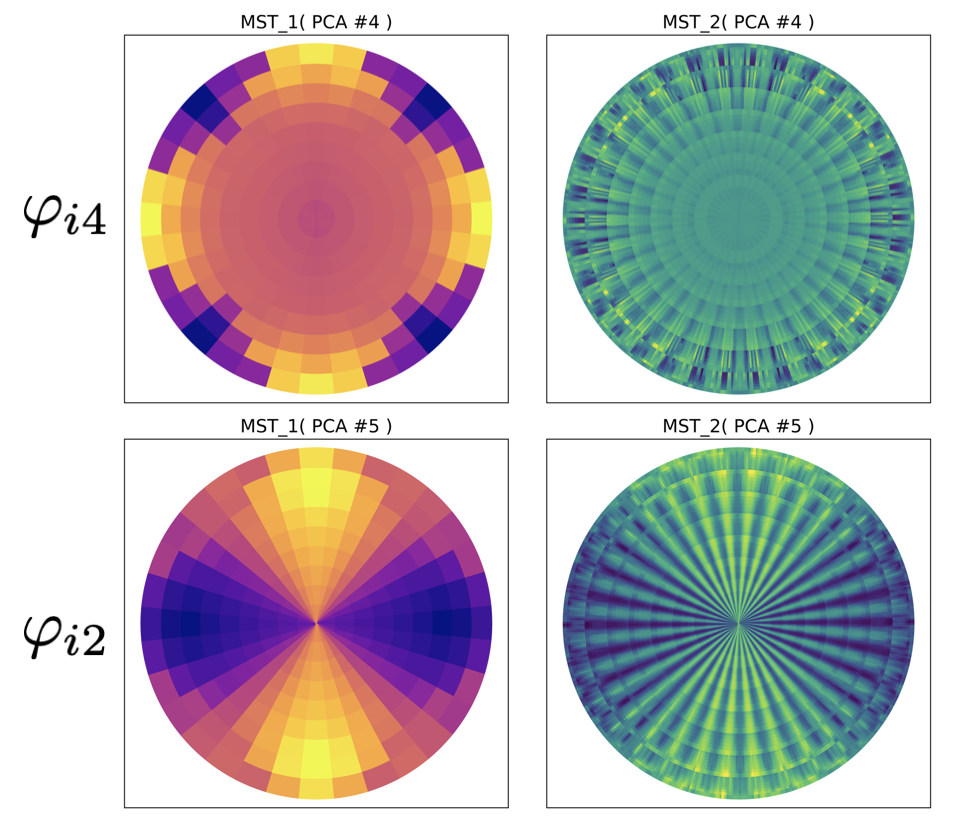}
\caption{\label{pca.self.organization} A closeup of two critical PCA vectors that illuminate the inverse cascade.  The radial axes are logarithmic in scale.  Shown are the first and second order MST for the $\varphi_{i4}$ (top) and $\varphi_{i2}$ (bottom) components.}
\end{figure}

\begin{figure}
\noindent\includegraphics[width=\columnwidth]{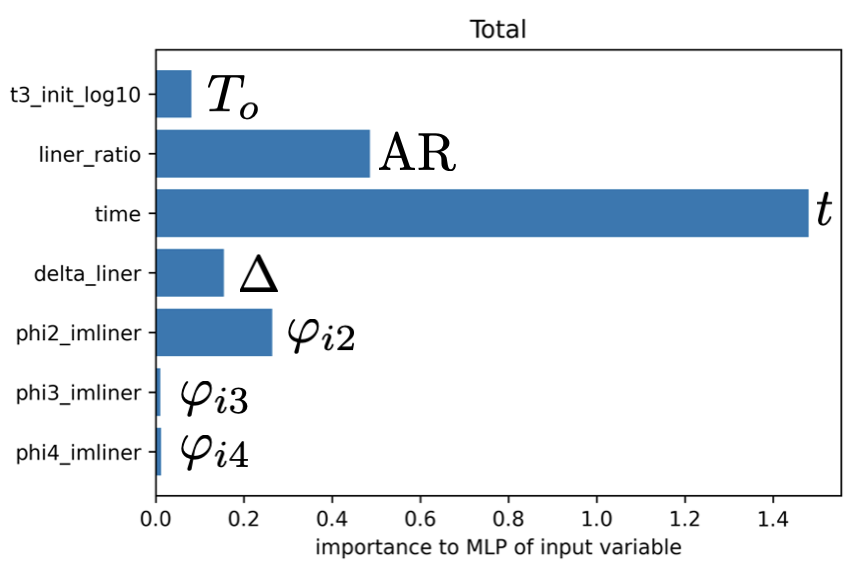}
\caption{\label{mlp.permutation.importance} MLP/NN permutation feature importance.}
\end{figure}

The results of the previous section give insight into the emergent behavior of this nonlinear system.  This behavior is exposed in the structure of two critical PCA vectors ($\varphi_{i4}$, the quadrupole moment with $m=4$, and $\varphi_{i2}$, the dipole moment with $m=2$). These PCA vectors are featured in Fig.~\ref{pca.self.organization}, where the scale (radial) axis is plotted on a logarithmic scale, and exposed in the permutation feature importance, as illustrated in Fig.~\ref{mlp.permutation.importance}.  Note that the $\varphi_{i4}$ PCA vector starts at the largest radii with a very clear quadrupole pattern in both the first order and the second order MST.  Due to the way that the second order MST is displayed, there will be two beats in every one of the 32 sectors.  This pattern disappears at the smaller radii as the plasma nears stagnation.  The opposite is true of the $\varphi_{i2}$ PCA vector.  The dipole pattern persists to the smallest radii.  Note that there is only one beat per sector for a dipole pattern.  There is even a strong overall dipole pattern on the second order MST.  The pattern seems to appear stronger as the radius gets smaller.  You can see this by focusing your attention on the outer three rings of the first order MST.  Note that as the $\varphi_{i4}$ PCA vector loses its structure, the structure of the $\varphi_{i2}$ PCA vector increases.  This is happening because there is an inverse cascade from the $\varphi_{i4}$ PCA vector to the the $\varphi_{i2}$ PCA vector, forming a self-organized dipole state.  This state persists as the plasma expands, post stagnation, to a larger radius.  This is further highlighted by the permutation feature importance, shown in Fig.~\ref{mlp.permutation.importance}.  It should be no surprise that time is the most important feature, followed by $\text{AR}$.  The liner aspect ratio changes the acceleration, which effects all facets of the evolution.  An interesting result is that the phase of the $m=2$ mode is the next most important feature.  On reflection, this is not surprising.  The initial phase of the $m=2$ mode, although a stochastic variable, is very quickly reinforced by the inverse cascade into this mode, and sets the phase of the resulting dipole mode.  The size of the perturbation does not matter as much, as demonstrated by $\Delta$ being less important than $\varphi_{i2}$, and $\varphi_{i3}$ and $\varphi_{i4}$ having little to no importance.  The preheat temperature $T_0$ only effects the evolution near stagnation, so its modest importance is expected.
%===============================%
%===============================%
%===============================%
%===============================%

\begin{figure}
\noindent\includegraphics[width=\columnwidth]{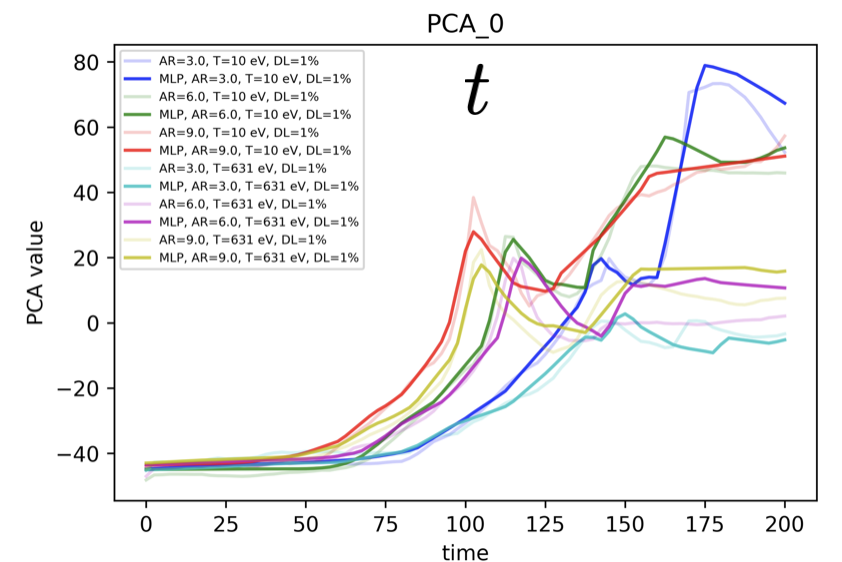}
\caption{\label{mlp.pca0} The time evolution of the first PCA vector component with respect to time.  The MLP/NN estimated values are the bold lines, the \texttt{GORGON} simulated values are the light lines in corresponding colors.  Shown are six cases for $T_0= 10 \, \text{eV}, 631 \, \text{eV}$ and $\text{AR}=3,6,9$.}
\end{figure}

Let us now focus our attention on the detail of the evolution being predicted by the MLP/NN by looking closely at the prediction of the first PCA vector component in Fig.~\ref{mlp.pca0}.  Remember that a MLP/NN with ReLU activation layers is a piecewise linear universal function approximator.  There are few tie points where the function is linear, but the number increases near the singularities in the mapping, where the slope is changing rapidly.  A couple of tie points are always put near stagnation.  The liner with the largest $\text{AR}$ has the largest acceleration and should converge first.  Indeed, the two simulations with $\text{AR} = 9$, shown in red and yellow, do converge first.  This time should be roughly independent of the preheat temperature $T_0$ (which it is), and the stagnation with the smaller $T_0=10 \, \text{eV}$ should have more compression than the one with $T_0=631 \, \text{eV}$ (which it does).  This feature is also present with the $\text{AR}=6$ stagnations (green and magenta), and the $\text{AR}=3$ stagnations (blue and cyan).  The MLP/NN captures this feature well.  %The trend continues with the $\text{AR}=6$ stagnations (green and magenta), and the $\text{AR}=3$ stagnations (blue and cyan).
%===============================%
%===============================%

There are several astonishing things that have happened in this MLDL workflow, that lead to the following questions.  Why did the dynamics reduce down to the evolution of a few PCA vector components?  Another way of asking this is: why are the dynamics constrained to a very low dimensional linear subspace of this complex (in the case of the WPH) Hilbert space?  What is the physical interpretation of the basis vectors that span this low-dimensional linear subspace?  Why are both the SVDs (inputs to outputs, and PCA projection to SVD projection) nearly diagonal?  Why were the $\log_{10}$ operations needed before and after the functional convolutional transformation (the MST and WPH)?  Why was the best MLP/NN architecture a encoder/decoder architecture, and what is the physical interpretation of the coordinates of the middle layer?  Why was it so easy for the MLP/NN to approximate the function with so few neurons?  The answer to these questions will be the subject of an upcoming paper, now in draft form.  Because the design of this workflow was not accidental nor the product of a large amount of experimentation (in fact, all the best design choices were the first thing that we tried, including the structure of the MLP/NN), we will present the answers to these questions, which you should view as hypotheses at this time.  The efficacy of this MLDL is tantalizing evidence supporting these hypotheses.  

Could the MST, if properly formulated, be a transformation to a complex ``renormalization'' space where the basis vectors are the solution to the Renormalization Group Equations (RGE)? (Remember that Renormalization is the study of how the physics changes as a function of scale, in our case $p$.  The solution to the RGEs, a coupled set of ODEs, gives the scaling exponents as a function of scale, where the fields and coupling constants scale as $\sim f_i^{\beta_i(p)}$, where the $\beta_i(p)$ are the solutions to the RGEs for field $f_i$.)  Is a natural logarithm needed to flatten this space?  Is there then a simple mapping to decoupled action-angle coordinates on the low-dimensional linear subspace, where the action and the angle are uniform circular functions of time?  After all, the RGEs are coupled ODEs of the logarithmic derivative. The complex $\ln(z)$ function, as a conformal mapping, takes polar coordinates about the origin to the cylinder where 
\begin{equation}
    \ln(z)= \ln |z| + \text{i} \, \arg(z),
\end{equation}
which flattens the space.  A PCA then would identify this linear subspace.  An SVD analysis of the fields and coupling constants would diagonally correlate to the basis vectors of this subspace.  The PCA vectors would be the solutions to the RGEs.  The number of important PCA vectors would therefore be equal to the number of fields and coupling constants.  The number of actions plus angles would be twice the number of PCA vectors.  Given that there are 6 fields (density, charge density, and 4 E\&M fields) and 4 coupling constants (charge of the electron and ion, and mass of the electron and ion, which were held constant) in MHD, the use of 7 PCA vectors and 15 nodes in the middle hidden layer is not surprising.  There are six fields and another adjoint basis vector for the resistivity, giving a total of seven.  The encoding in the middle layer needs to have a node for the action and the angle for each field and another for time, giving a total of 15.   The motion on this low dimensional linear subspace would be geodesic motion determined by an analytic function that the MLP/NN is approximating.  It is interesting to note that knowing the topology of this low-dimensional complex space is equivalent to knowing the analytic function (it is the solution to Laplace's equation).  For a simple topology, this would be knowing the location and order of the poles and zeros.  It would be very easy for a MLP/NN to approximate this function since it is a solution to Laplace's equation.  It would need few tie points away from the poles and zeros because the space would be flat.

These hypotheses, if true, lead to some interesting applications of this MLDL workflow.  For example, the predictions will extrapolate well, as long as the extrapolation is made going away from the poles and zeros of the topology.  This can be experimentally tested.  First, make a prediction using the MLDL predictor of scaling into into a new regime, then do the experiment.  If the prediction is better than expected, then the extrapolation is away from the poles and zeros.  If the prediction is worse than expected, then the extrapolation is towards new poles and zeros.  The theoretical model needs to have additional physics added to it.  In fact, the MLDL workflow, if augmented with the experimental points and the deletion of the theoretical (computer simulation) in this extrapolation regime, will determine what topology needs to be added.  This process is one of topological discovery or causal analysis.

Another way of looking at the MLDL workflow presented and implemented in this paper is a redefinition of the MST/WPH to
\begin{equation}
    S_m[f(x)](p,x) \equiv \phi_{px} \star \left( \prod_{k=1}^m{\text{i} \ln \, R_0 \, \psi_{p_k} \star} \right) \text{i} \ln \, R_0 \, f(x),
\end{equation}
where 
\begin{equation}
    R_0(z) \equiv z + \text{e}^{\text{i} \, \arg(z)}.
\end{equation}
It should be noted that this transformation is no longer stationary since the Father Wavelet only averages over as large of a patch as it has to do.  Nothing prevents this partition of unity from being summed over a larger domain in $x$, if the process is stationary over that domain.  This transformation is complex from the beginning to the end.  The real part is the $\ln(\text{mod})$ and the imaginary part is the $\text{arg}$.  Not only does $\ln(R_0(z))$ conformally flatten the space onto the cylinder, it now (with the addition of $R_0$) exponentially (for large deviation), then logarithmically (for small deviations) converges to the origin $z=0$.  The connection to the MST/WPH work can be seen by examining the limiting behavior of the $\ln(R_0(z))$ mapping
\begin{equation}
\begin{split}
    \ln(R_0(z)) &\xrightarrow[|z| \to 0]{} z \\
           &\xrightarrow[|z| \to \infty]{} \ln|z| \quad \text{(this work)}\\
           &\xrightarrow[|z| \to \infty]{} \ln|z| \, \text{e}^{\text{i} \, \arg(z) / \ln|z|}\\
           &\sim \sum_{k=0}^n{|z| \, \text{e}^{\text{i} \, k \, \arg{(z)}}} \quad \text{(that is, WPH)}\\
           &\xrightarrow[n = 0]{} |z| \quad \text{(that is, MST)}.
\end{split}
\end{equation}
The characteristic of the fixed point at the origin is the statement of the first limit. The second limit is effectively what has been used in this paper.  In the third limit (which keeps the small $\arg$ imaginary term), the $\ln$ chirps the pulse, generating the harmonics that are explicitly generated in the fourth line by the WPH.  The conventional MST is just using the first term in that series.

\begin{figure}
%\noindent\includegraphics[width=\columnwidth]{figures/mldl_workflows.png}
%%%%%
\textbf{Composite approach} \\ [0.3cm]
\begin{minipage}{0.08\textwidth}
\vspace{0.25cm}
\[
\displaystyle
\begin{bmatrix}
\tilde{A} \\
C
\end{bmatrix}
\]
\end{minipage}
\hspace{1.5cm}
\begin{minipage}{0.1\textwidth}
\quad $\tilde{D}(\tilde{A}, C)$ \\ [-0.2cm]
\setlength{\unitlength}{0.2cm}
\begin{picture}(0,0)
    \put(-15, -2.5){\vector(1,0){10}}
    \put(-14.5,-4.3){($n_{A} \! + \! n_{C})$-D}
    \put(7.5, -2.5){\vector(1,0){5.5}}
    \put(8.2,-4.3){$n_{D}$-D}
\end{picture}
\fcolorbox{black}{blue!10}{
\begin{minipage}{\textwidth}
\vspace{0.2cm}
MLP/NN
\vspace{0.2cm}
\end{minipage}
}
\end{minipage}
\hspace{1.1cm}
\begin{minipage}{0.08\textwidth}
\[
\displaystyle
\begin{bmatrix}
\tilde{D}
\end{bmatrix}
\]
\end{minipage}
%%%%%
\\ [0.4cm]
\textbf{Decomposed approach} \\ [0.3cm]
\begin{minipage}{0.04\textwidth}
\[
\displaystyle
\begin{bmatrix}
\tilde{A}
\end{bmatrix}
\]
\end{minipage}
\hspace{0.85cm}
\begin{minipage}{0.1\textwidth}
\quad $\tilde{z}_{i}(\tilde{A})$ \\ %[-0.2cm]
\fcolorbox{black}{blue!10}{
\begin{minipage}{\textwidth}
\vspace{0.2cm}
MLP/NN
\vspace{0.2cm}
\end{minipage}
}
\end{minipage}
\hspace{1.1cm}
\begin{minipage}{0.04\textwidth}
\[
\displaystyle
\begin{bmatrix}
\tilde{z}_{i}
\end{bmatrix}
\]
\end{minipage}
\\ 
\setlength{\unitlength}{0.2cm}
\begin{picture}(15,0)
    \put(-3,2.7){\vector(1,0){5}}
    \put(-2.5,0.9){$n_{A}$-D}
    \put(14,2.7){\vector(1,0){4}}
    \put(14.5,0.9){$n$-D}
    \put(20,1.2){\line(0,-1){2}}
    \put(20,-0.8){\line(-1,0){28.5}}
    \put(-8.5,-0.8){\vector(0,-1){4.8}}
\end{picture}
\\ [0.2cm]
Dynamic Mapper \\ [0.2cm]
\hspace{1.5cm} $\tilde{z}(\tilde{z}_{i}, C; t)$ \\ [-0.1cm]
\begin{minipage}{0.04\textwidth}
\vspace{-0.28cm}
\[
\displaystyle
\begin{bmatrix}
\tilde{z}_{i} \\
C \\
t
\end{bmatrix}
\]
\end{minipage}
\hspace{2.3cm}
\begin{minipage}{0.1\textwidth}
\setlength{\unitlength}{0.2cm}
\begin{picture}(0,0)
    \multiput(-21,-6.5)(0,1.4){7}{\line(0,1){0.8}}
    \multiput(-21,-6.5)(1.4,0){26}{\line(1,0){0.8}}
    \multiput(-21,2.7)(1.4,0){26}{\line(1,0){0.8}}
    \multiput(14.8,-6.5)(0,1.4){7}{\line(0,1){0.8}}
    \put(-17, -2.5){\vector(1,0){12}}
    \put(-16.5,-4.3){($n_{A} \! + \! n_{C} \! + \! 1)$-D}
    \put(7.5, -2.5){\vector(1,0){4}}
    \put(8,-4.3){$n$-D}
    \put(13,-4){\line(0,-1){3.5}}
    \put(13,-7.5){\line(-1,0){28.5}}
    \put(-15.5,-7.5){\vector(0,-1){3.5}}
    %
    %\put(-3,2.7){\vector(1,0){5}}
    %\put(-2.5,0.9){$n_{A}$-D}
    %\put(14,2.7){\vector(1,0){4}}
    %\put(14.5,0.9){$n$-D}
    \put(-13.5, -12.5){\vector(1,0){4}}
    \put(-13, -14.3){$n$-D}
    \put(2.5, -12.5){\vector(1,0){5}}
    \put(3, -14.3){$n_{D}$-D}
\end{picture}
\fcolorbox{black}{blue!10}{
\begin{minipage}{\textwidth}
\vspace{0.2cm}
MLP/NN
\vspace{0.2cm}
\end{minipage}
}
\end{minipage}
\hspace{1.1cm}
\begin{minipage}{0.04\textwidth}
\vspace{-0.35cm}
\[
\displaystyle
\begin{bmatrix}
\tilde{z}
\end{bmatrix}
\]
\end{minipage}
\\ [0.5cm]
\begin{minipage}{0.04\textwidth}
\[
\displaystyle
\begin{bmatrix}
\tilde{z}
\end{bmatrix}
\]
\end{minipage}
\hspace{0.65cm}
\begin{minipage}{0.1\textwidth}
\quad $\tilde{D}(\tilde{z})$ \\ %[-0.2cm]
\fcolorbox{black}{blue!10}{
\begin{minipage}{\textwidth}
\vspace{0.2cm}
MLP/NN
\vspace{0.2cm}
\end{minipage}
}
\end{minipage}
\hspace{1.3cm}
\begin{minipage}{0.04\textwidth}
\[
\displaystyle
\begin{bmatrix}
\tilde{D}
\end{bmatrix}
\]
\end{minipage}
%
%%%%%
\\ [0.4cm]
\textbf{Current work} \\ [0.3cm]
\begin{minipage}{0.08\textwidth}
\vspace{0.25cm}
\[
\displaystyle
\begin{bmatrix}
\tilde{A} \\
t
\end{bmatrix}
\]
\end{minipage}
\hspace{1.35cm}
\begin{minipage}{0.1\textwidth}
\quad $\tilde{z}(\tilde{A}, t)$ \\ [-0.2cm]
\setlength{\unitlength}{0.2cm}
\begin{picture}(0,0)
    \put(-14, -2.5){\vector(1,0){9}}
    \put(-13.5,-4.3){($n_{A} \! + \! 1$)-D}
    \put(7.5, -2.5){\vector(1,0){5.5}}
    \put(8.2,-4.3){$n_{D}$-D}
\end{picture}
\fcolorbox{black}{blue!10}{
\begin{minipage}{\textwidth}
\vspace{0.2cm}
MLP/NN
\vspace{0.2cm}
\end{minipage}
}
\end{minipage}
\hspace{1.1cm}
\begin{minipage}{0.08\textwidth}
\[
\displaystyle
\begin{bmatrix}
\tilde{z}
\end{bmatrix}
\]
\end{minipage}
\caption{\label{mldl.workflows} Three different variations of the MLDL workflows: (top) a composite workfow, (middle) a decomposed approach, and (bottom) the hybrid case presented in this paper.}
\end{figure}
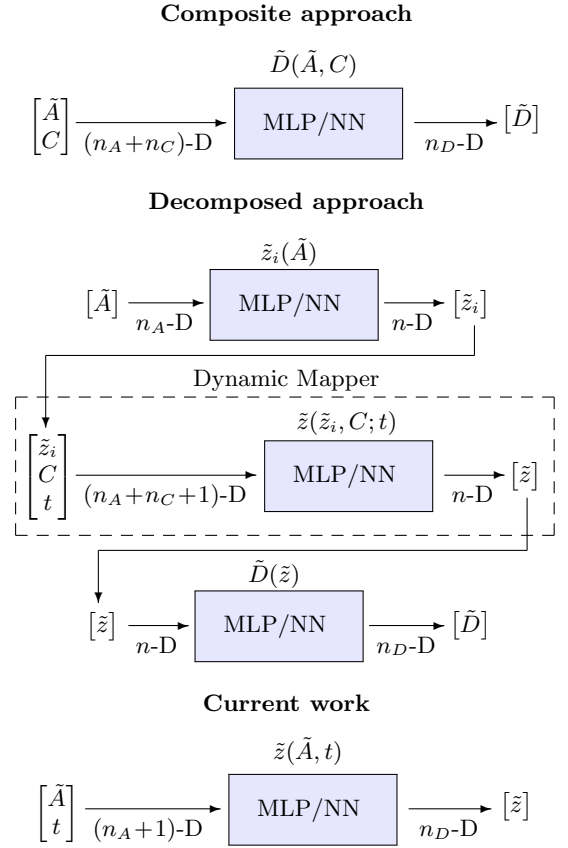

This MLDL architecture can be modified to make it a sequential, modular approach as shown in Fig.~\ref{mldl.workflows}.  In general, one starts with initial conditions $\Tilde{A}$ and coupling constants $C$, does a computer simulation of the process, followed by a computer simulation of the diagnostics, to predict the diagnostic response $\Tilde{D}$.  Here the tilde signifies the $\text{PCA}(\ln(\text{MST/WPH}(\ln)))$ of the field quantity.  The composite approach can be taken, where there is one MLP/NN, $\Tilde{D}(\Tilde{A},C)$.  There is a more flexible decomposed approach where the approximation is separated into three approximators.  The first is a mapping of the initial conditions to the initial fields $\Tilde{z}_i(\Tilde{A})$.  This is followed by a general dynamic mapper of the initial fields, coupling constants and time to the fields $\Tilde{z}(\Tilde{z}_i,C;t)$.  Finally, there is a mapping from the fields to the diagnostics, $\Tilde{D}(\Tilde{z})$.  The current work is a restricted hybrid of initial conditions and time to the fields $\Tilde{z}(\Tilde{A};t)$.
%===============================%
%===============================%

This work needs to be extended into 3D to see if the inverse cascade persists and what the characteristics of the 3D self-organized state are.  It is well-known that the inverse cascade in 2D fluid flow is caused by the topological invariant of vorticity.   For the case of the magnetized plasmas encountered in MagLIF, there is still a topological invariant: helicity \citep{perez2009role, glinsky2019helicity, taylor1986relaxation}.  There is experimental evidence of helical stagnations showing the dipole structure in the plane perpendicular to the axis \citep{Awe2013}.  In addition, there is also an axial sausage mode.  The result is a double helical structure that looks like a DNA atom.

There are obvious improvements that also need to be made to the MST/WPH.  It needs to be made non-stationary, such that it is not translationally invariant.  The complex natural logarithm function needs to be made an integral part of it.  The orthogonal local wavelet basis needs to be a carefully constructed partition of unity, so that a fast inverse transformation can be constructed.  Given this basis, a resolution independent, physical display needs to be constructed.

We end with a summary of what this research has shown or given indications may be true.  It has demonstrated a fast, high fidelity surrogate for resistive MHD.  This surrogate is $10^7$ times faster than conventional computational prediction.  It is based on a simple, fast to train, physics-based machine learning.  It gives field to field correlation, physically interpretable results, and meaningful graphical displays.  It has the potential to give fundamental insights into nonlinear dynamics, physical kinetics, quantization and second quantization, renormalization, and the topology of dynamics.  This surrogate will either extrapolate well or give insights into additional causality.  Finally, from the practical MagLIF physics perspective, it has shown an emergent behavior of 2D MagLIF implosions into a self organized dipole state.

\begin{acknowledgments}
We would like to thank St\'ephane Mallat, Joan Bruna, Joakim Anden, Sixin Zhang, John Field, Pat Knapp, Nat Trask, and Eric Cyr for many useful discussions.  We are grateful to Chris Jennings for providing the version of \texttt{GORGON} used to generate the training samples, and to Lawrence Livermore National Laboratory and Marty Marinak for providing the high performance computational facilities.  Francis Ogoke and Amir Barati Farimani gave much useful advice with the MLDL aspects of this research.  Thanks is given to CSIRO for supporting some of the early work (MEG) through their Science Leaders Program, the Institut des Hautes Etudes Scientifique (IHES) for hosting a stay where much was learned (MEG) about the MST, and the Santa Fe Institute for hosting a stay (MEG) where understanding of the MST was developed.  Finally, thanks is given to the University of Western Australia, John Hedditch, and Ian MacArthur for their help in understanding many of the finer points of Quantum Field Theory.  Most of this research was funded by the Sandia National Laboratories' Laboratory Directed Research and Development (LDRD) program.  Sandia National Laboratories is a multimission laboratory managed and operated by National Technology and Engineering Solutions of Sandia LLC (NTESS), a wholly owned subsidiary of Honeywell International Inc., for the U.S. Department of Energy's National Nuclear Security Administration (NNSA) under contract DE-NA0003525.  This paper describes objective technical results and analysis. Any subjective views or opinions that might be expressed in the paper do not necessarily represent the views of the U.S. Department of Energy or the United States Government.  Sandia National Labortories report number SAND2023-12134 O.  The data that support the findings of this study are available from the corresponding author upon reasonable request.  Much of the software which produced the results of this paper is available on \href{https://github.com/glinsky007/mstmhd}{Github} under an MIT open source license.
\end{acknowledgments}

\bibliography{mst_mhd_refs}% Produces the bibliography via BibTeX.

\end{document}